\documentclass[11pt, a4paper, oneside, reqno]{amsart}
\usepackage{comment}
\usepackage[usenames, dvipsnames]{color}
\definecolor{darkblue}{rgb}{0.0, 0.0, 0.45}
\definecolor{lightblue}{RGB}{240,248,255}
\definecolor{lightblue2}{rgb}{0.68, 0.85, 0.9}
\definecolor{lightcyan}{rgb}{0.88, 1.0, 1.0}
\definecolor{palepink}{rgb}{0.98, 0.85, 0.87}

\usepackage[colorlinks	= true,
raiselinks	= true,
linkcolor	= darkblue, 
citecolor	= Mahogany,
urlcolor	= ForestGreen,
pdfauthor	= {Peyman Mohajerin Esfahani},
pdftitle	= {},
pdfkeywords	= {},
pdfsubject	= {},
plainpages	= false]{hyperref}

\usepackage{dsfont,amssymb,amsmath,enumitem} 
\usepackage{amsfonts,dsfont,mathtools, mathrsfs,amsthm} 
\usepackage[amssymb, thickqspace]{SIunits}
\usepackage{fancyhdr,mdframed,nicefrac}

\usepackage[norelsize, ruled, vlined]{algorithm2e}
\usepackage{algcompatible}

\usepackage{epsfig}
\usepackage{graphicx}
\usepackage{float}
\usepackage{caption}
\usepackage{subcaption}

\usepackage{courier}

\usepackage{multirow}
\usepackage{bigstrut}

\allowdisplaybreaks
\date{\today}
\makeatletter
\def\@settitle{\begin{center}%
		\baselineskip14\p@\relax
		\normalfont\LARGE\scshape\bfseries
		\@title
	\end{center}%
}

\def\@setauthors{%
  \begingroup
  \def\thanks{\protect\thanks@warning}%
  \trivlist
  \centering\footnotesize \@topsep30\p@\relax
  \advance\@topsep by -\baselineskip
  \item\relax
  \author@andify\authors
  \def\\{\protect\linebreak}%
  \authors%
  \ifx\@empty\contribs
  \else
    ,\penalty-3 \space \@setcontribs
    \@closetoccontribs
  \fi
  \endtrivlist
  \endgroup
}

\makeatother
\makeatletter

\def\subsection{\@startsection{subsection}{2}%
	\z@{.5\linespacing\@plus.7\linespacing}{.5\linespacing}%
	{\normalfont\large\bfseries}}

\def\subsubsection{\@startsection{subsubsection}{3}%
	\z@{.5\linespacing\@plus.7\linespacing}{.5\linespacing}%
	{\normalfont\itshape}}

\usepackage{multirow}
\usepackage{bigstrut}
\usepackage{wrapfig}
\usepackage{caption}

\renewcommand{\geq}{\geqslant}
\renewcommand{\ge}{\geqslant}
\renewcommand{\le}{\leqslant}
\renewcommand{\leq}{\leqslant}

\DeclareSymbolFont{symbolsC}{U}{pxsyc}{m}{n}

\usepackage{bm}
\usepackage{amsthm}
\DeclareMathOperator*{\argminA}{arg\,min}

\usepackage{amssymb}
\usepackage{latexsym}
\usepackage{mathrsfs}  
\usepackage{bbm}
\usepackage{amsthm}

\usepackage{url}
\usepackage{color}

\usepackage{algorithm2e}
\usepackage{amsmath,bm}
\usepackage{subcaption}
\usepackage[makeroom]{cancel}

\newtheorem{prop}{Proposition}
\usepackage{float}
\usepackage{lipsum}
\usepackage{dsfont}
\usepackage{rotating}
\usepackage{mathtools}
\usepackage{relsize}

\title[Wasserstein-penalized Entropy closure]{
Wasserstein-penalized Entropy closure:\\ A use case for stochastic particle methods
}
 \author{Mohsen Sadr, Nicolas G. Hadjiconstantinou, and M. Hossein Gorji}
 \thanks{Corresponding author: Mohsen Sadr}
 \thanks {Emails: mohsen.sadr@psi.ch, ngh@mit.edu, mohammadhossein.gorji@empa.ch}
 \thanks{Mohsen Sadr: Department of Mechanical Engineering, MIT, Cambridge, MA 02139, USA and Paul Scherrer Institute, Forschungsstrasse 111, CH-5232 Villigen, Switzerland. Nicolas G. Hadjiconstantinou: Department of Mechanical Engineering, MIT, Cambridge, MA 02139, USA. M. Hossein Gorji: Laboratory of Multiscale Studies in Building Physics, Empa, Dübendorf, Switzerland.}
 
\begin{document}
\maketitle

\begin{abstract}
We introduce a framework for generating samples of a distribution given a finite number of its moments, targeted to particle-based solutions of kinetic equations and rarefied gas flow simulations. Our model, referred to as the Wasserstein-Entropy distribution (WE), couples a physically-motivated Wasserstein penalty term to the traditional maximum-entropy distribution (MED) functions, which serves to regularize the latter. The penalty term becomes negligible near the local equilibrium, reducing the proposed model to the MED, known to reproduce the hydrodynamic limit. However, in contrast to the standard MED, the proposed WE closure can cover the entire physically realizable moment space, including the so-called Junk line. We also propose an efficient Monte Carlo algorithm for generating samples of the unknown distribution which is expected to outperform traditional non-linear optimization approaches used to solve the MED problem. Numerical tests demonstrate that, given moments up to the heat flux---that is equivalent to the information contained in the Chapman-Enskog distribution---the proposed methodology provides a reliable closure in the collision-dominated and early transition regime. Applications to larger rarefaction demand information from higher-order moments, which can be incorporated within the proposed closure. 
\end{abstract}%


\section{Introduction}
\label{sec:introduction}

\noindent  Kinetic theory provides a probabilistic description of mesoscale transport processes well beyond the continuum limit. Although it is an approximation of molecular transport in the limit of a dilute system of particles, it has found application in a wide variety of fields, such as rarefied gas flow \cite{Cercignani,cercignani2000rarefied,sone2007molecular}, evaporation/condensation phenomena \cite{frezzotti2005mean,sadr2021fokker}, solid-state heat transfer as mediated by phonons \cite{chen2005nanoscale,peraud2011efficient,peraud2012alternative} and plasma dynamics \cite{Chapman1953,Hirschfelder1963,klimontovich1975kinetic}. 
\\ \ \\
Kinetic descriptions are invaluable for describing the transition from collision-dominated (diffusive) behavior for $\mathrm{Kn}\ll 1$ to collisionless (ballistic) behavior for $\mathrm{Kn}\gg 1$ \cite{hadjiconstantinou2006limits}, where $\mathrm{Kn}=\lambda/L$ denotes the Knudsen number. Recent applications are typically related to small-scale science and engineering where the carrier mean free path,  $\lambda$, becomes appreciable compared to or even larger than the characteristic transport lengthscale, $L$ \cite{karniadakis2005microflows}. Although kinetic-theory descriptions are able to capture the continuum behavior, and in fact inform its constitutive behavior \cite{Chapman1953,Cercignani,sone2007molecular}, computational methods based on kinetic theory, such as direct simulation Monte Carlo (DSMC) \cite{Bird} and related variants \cite{nanbu1980direct} including variance-reduced methods \cite{homolle2007low,radtke2011low,al2010excursion,sadr2023variance}, direct solution methods 
\cite{broadwell_1964,oblapenko2020velocity}, finite-difference methods \cite{ohwada1993structure} and spectral methods \cite{Pareschi2000,gamba2017fast}, become stiff in this limit \cite{pareschi2001time,trazzi2009adaptive}.  
\\ \ \\
This stiffness has motivated the introduction of a number of approximate models, including the Fokker-Planck model \cite{Jenny2010,Gorji2011}, the Lattice Boltzmann method \cite{succi2001lattice},  Bhatnagar-Gross-Krook (BGK) based unified gas-kinetic models \cite{xu2010unified,liu2020unified}, the jump-diffusion approximation \cite{mies2023efficient} and moment methods \cite{torrilhon2016modeling}.
At the same time, a number of computational multiscale approaches for bridging this length- and timescale gap have been proposed. For example, hybrid methods which use the kinetic solution method only in the region where it is needed and use a continuum solution method in the remainder of the computational domain have been developed \cite{garcia1999adaptive,tiwari2009particle,di2016lattice,sadr2021coupling}. Here we also mention asymptotically preserving methods \cite{Toscani,Lemou} and deviational variance-reduction methods \cite{radtke2009variance,peraud2011efficient,peraud2012alternative}. The latter use algebraic decomposition \cite{radtke2013efficient}---in contrast to physical-domain decomposition used by hybrid methods---to seamlessly transition between an analytical description of the distribution function and a particle-based numerical solution; as expected, the closer the analytical description to the actual distribution, the larger the computational savings \cite{radtke2009variance,radtke2013efficient}. In particular, methods which make use of the local equilibrium distribution as a control variate, although more complex \cite{radtke2009variance},  perform significantly better than traditional particle methods as the continuum limit is approached, since the analytical description becomes an increasingly better approximation of the gas state, thus minimizing the amount of computation required \cite{radtke2009variance,sadr2023variance}.
\\ \ \\
The importance of bridging continuum and atomistic representations of transport extends well beyond the realm of kinetic theory and dilute gases. Beyond hybrid methods coupling molecular dynamics simulations to Navier-Stokes simulations for dense fluid problems \cite{hadjiconstantinou1997heterogeneous,hadjiconstantinou1999hybrid,werder2005hybrid}, more recent work has led to a wide range of methods that attempt to efficiently combine the fidelity of atomistic methods with the computational efficiency of continuum solution methods using extensions of continuum analyses and matching techniques (see, for example, \cite{kevrekidis2003equation,gear2003projective} and references therein). Such methods include projective time integration \cite{gear2003projective,kevrekidis2004equation}, patch dynamics \cite{gear2003gap,samaey2006patch}, the Heterogeneous Multiscale Method \cite{weinan2003heterogeneous}, as well as methods for finding fixed points \cite{kevrekidis2004equation}. Although more general than some of the techniques reviewed above, in the sense that they do not use any information on the governing equation at the microscopic level, these techniques rely on accurate methods of matching between the continuum and atomistic description. While passing information about the atomistic field to the continuum description can be achieved straightforwardly by a process of averaging or "restriction" \cite{kevrekidis2003equation}, the reverse process, namely initializing or imposing boundary conditions on an atomistic simulation from information obtained from a continuum solution, is significantly more challenging and is the main reason for these methods not reaching their full potential. Specifically, the challenge lies in the fact that {\it in general}, knowledge of the continuum solution, which corresponds to the first few moments of the particle distribution function, is insufficient to completely describe the molecular state and thus the complete distribution function. 
\\ \ \\
Beyond the applications to multiscale computation discussed above, the need for determining an unknown distribution function or generating samples from it given some of its moments is important in many areas of physics.
Focusing on kinetic theory computations, applications can be found in problems involving variance reduction \cite{al2010excursion,moment-guided,sadr2023variance,sadr2023FP}, to accelerating convergence to or direct solution for steady states \cite{al2007acceleration}, to improving accuracy by ensuring moment conservation.
\\ \ \\
In this work we tackle this fundamental problem in the case of kinetic theory models; namely, given a small number of moments of the molecular-velocity distribution function, we propose an efficient Monte Carlo methodology for creating samples from a distribution function that differs from the true particle distribution in a least bias sense. Using the known moments as constraints, we use the least bias principle in information theory to devise a  closure by minimizing the Shannon entropy as well as the Wasserstein distance from the local Maxwellian distribution. The proposed approach is an extension of the maximum entropy approach in the sense that the additional Wasserstein penalty term has a regulating effect on the maximum entropy formulation, allowing existence for all realizable moments while maintaining the convexity of the underlying optimization problem.  We recall that the standard Maximum entropy distribution (MED) function suffers from unrealizablity (degeneracy) for some physically realizable moments, e.g. on the Junk-line \cite{junk1998domain}, also manifested as a high condition number in the underlying optimization problem near the limit of physical realizability \cite{abramov2007improved,alldredge2014adaptive,alldredge2019regularized}. At the same time, the Wasserstein term allows fast convergence to the MED for near-equilibrium moment problems, as its contribution vanishes to the first order.
\\ \ \\
 The remainder of this paper is organized as follows. In Sec. \ref{sec:WE_MainIdea}, we motivate the main idea behind the proposed approach and its relation to kinetic theory. In Sec. \ref{sec:MomentSystemLimit}, we formulate the closure problem and show its convergence to the Navier-Stokes-Fourier system using a Chapman-Enskog-type expansion. In Sec. \ref{sec:StochRep}, we devise a stochastic process for generating samples from the proposed closure. In Sec.  \ref{sec:SolAlg_SanityCheck}, we validate the accuracy and robustness of the proposed solution algorithm using a number of numerical tests, while in Sec. \ref{sec:results} we use the proposed closure in test problems involving DSMC calculations. Finally, we discuss our conclusions and outlook in Sec. \ref{sec:conclusion}.
\section{Wasserstein-Penalized Entropy Closure}
\label{sec:WE_MainIdea}

\subsection{Main Idea}

\noindent We seek a solution for the following closure problem: infer a probability density $\bar{f}(v)$ on $\mathbb{R}^m$ from a finite set of its moments.
In general, this inverse problem is ill-posed, and hence further assumptions/regularizations need to be introduced. Overall there exist two categories of algorithms. One focuses on expanding the unknown density with respect to some basis functions, e.g. in the Grad method a Hilbert expansion of the distribution function in the Hermite polynomials in the pursuit of relations describing the dependence of high-order moments of the distribution on its low-order moments \cite{grad1958principles,torrilhon2016modeling}. The other focuses on minimizing some cost functional subject to the moment constraints (see e.g. \cite{henrion2022graph}). In particular, in MED formulations the optimization is based on entropy minimization{\footnote{The physical entropy $- \int_{\mathbb{R}^m} f (\log f-1) {\textrm {dv}}$ is a concave function of $f$, whose maximum coincides with the minimum of the convex mathematical entropy $\int_{\mathbb{R}^m} f (\log f - 1) {\textrm {dv}}$, also known as the kinetic entropy \cite{hauck2008convex}. In this work, we will be using the term entropy to refer to the latter.}} \cite{levermore1996moment}. Both categories might suffer from the lack of well-defined solutions for an arbitrary physical moment set. \\ \ \\
While conventional approaches directly postulate a functional form for $\bar{f}$, in this work we adopt an alternative path.
In particular, our strategy is to infer a joint probability density $\pi(v,w)$ on $\mathbb{R}^{2m}$ such that its marginal 
\begin{eqnarray}
f(v)&=&\int_{\mathbb{R}^m} \pi(v,w) \ \text{dw}
\end{eqnarray}
gives a solution to the closure problem (and hence an approximation of $\bar{f}$), whereas the other marginal 
\begin{eqnarray}
g(w)&=&\int_{\mathbb{R}^m} \pi(v,w) \ \text{dv} 
\label{eq:dist_w}
\end{eqnarray}
is linked to a known density $\bar{g}$, which serves as a means of introducing some prior knowledge (e.g. a nearby equilibrium state, a prior approximation, etc). 
Introducing the latter density allows us to control the distance of the inferred density from the known one in a suitable metric, and hence bring in further regularization for the unknown distribution. Throughout this work, we employ non-italic font to denote the infinitesimal volumes in the integrals, e.g. $\textrm{dv}$ and $\textrm{dw}$ stand for infinitesimal volumes around $v$ and $w$, respectively.
\\ \ \\
In what follows, we focus on two fundamental statistical concepts as guiding principles.
Motivated by the principle of the least action, minimizing the transport cost between the two marginals
\begin{eqnarray}
\mathcal{W}(\pi)&=\mathop{\mathlarger{\int}}_{\mathbb{R}^m \times \mathbb{R}^m} c(v,w)\ \pi(v,w) \ \text{d} v\text{d} w
\end{eqnarray}
can serve as a metric which controls the transport cost between $f$ and $g$ weighted with the cost function $c(v,w)$. We employ $c(v,w)=C_0\lvert v-w\rvert^p$, with $|\cdot |$ denoting the usual Euclidean norm in $\mathbb{R}^m$. The exponent $p\ge 2$ depends on the structure of the constraints, while the constant $C_0>0$ acts as a normalization factor (both will be fixed later).
\\ \ \\
Furthermore, we would like to incorporate the physical maximum entropy principle in the closure model by introducing an additional loss functional
\begin{eqnarray}
\mathcal{H}(\pi)&=&\int_{\mathbb{R}^m \times \mathbb{R}^m}  \bigg(\log(\pi(v,w))-1\bigg)\  \pi(v,w) \ \text{d} \text{v}\text{d} \text{w}.
\end{eqnarray}
Minimizing the entropy endows $\pi$ with favourable statistical features, such as the least bias property \cite{jaynes2003probability}. The extremum of the resulting convex functional
\begin{eqnarray}
\mathcal{L}_{\alpha}(\pi)&=&\alpha \mathcal{W}(\pi)+\mathcal{H}(\pi) \ \ \ \ \ \textrm{for} \ \ \ \ \alpha>0\ ,
\end{eqnarray}
referred to as the Wasserstein-penalized Entropy functional (WE),   represents our solution to the closure problem. The combination of penalty terms therein allows us to obtain a closed-form solution to the optimal $\pi$, while preventing regularity issues arising from isolated minimization of $\mathcal{H}$.
\\ \ \\
The proposed optimization can be seen as a generalization of two limiting cases of $\mathcal{L}_\alpha$. For $\alpha \to \infty$, we get a loss functional  close to the Sinkhorn distance (where the Kullback-Leibler divergence is used instead of the Shannon entropy $\mathcal{H}$). The Sinkhorn distance has been employed to accelerate the computation of optimal transport problems  \cite{genevay2016stochastic,cuturi2013sinkhorn,genevay2019sample}. The other limit, $\alpha\to 0$, gives the celebrated MED \cite{jaynes1957information,levermore1996moment,alldredge2012high,mcdonald2013affordable}.
The merit of the proposed loss $\mathcal{L}_{\mathcal{\alpha}}(\pi)$ lies in the fact that, by the proper choice of the exponent $p$, the resulting optimization construct prevents the degeneracy of MED, while allowing for efficient solution algorithms, as will be demonstrated in the rest of the manuscript. 
\subsection{Kinetic Context}
\noindent In the framework of gas kinetic systems, we usually deal with the three-dimensional  velocities $v,w\in \mathbb{R}^3$. Moreover, we are often interested in the dynamics of the single-particle distribution function at time $t$ and physical space $x\in \Omega\subseteq\mathbb{R}^3$. Therefore, the corresponding probability densities are evaluated conditioned on a given $x$ and $t$, i.e. $f( v| x, t)$. In the interest of simplicity, we drop the conditional on space $x$ and time $t$ and write $f(v)$, unless its dependence is needed for the analysis. Also, for convenience, we consider a scaling of $f(v)$, such that it returns the gas density $\rho(x,t)$, once integrated with respect to $v$.
\\ \ \\
Let $\mathbb{M}^{v}$ and $\mathbb{M}^w$ be linear sub-spaces of some polynomials in $v$ and $w$, respectively, each with the dimension $n$. Suppose $H^v(v)$ and $H^w(w)$ are the corresponding basis functions in $\mathbb{M}^{v}$ and $\mathbb{M}^{w}$, respectively. Take $H_i^v(v)\in H^v(v)$ and $H_i^w(w)\in H^w(w)$ to be the $i$-th elements of those polynomial basis functions. In this work, we use the subscript $(.)_i$ to denote $i$-th component of a vector. For convenience, we assume $H_i^v(v)=H_i^w(w),\forall v=w$, and suppose $H$ denotes the column vector representing their union $H=[H_1^v,...,H_n^v,H_1^w,...,H_n^w]^T$. Finally, the integrals with respect to the measure associated with a probability density $h$ are denoted by $\langle \ \cdot \ \rangle_h$, e.g.
\begin{eqnarray}
\langle \phi \rangle_f&=&\int_{\mathbb{R}^3} \phi(v) f(v) \ \textrm{dv} .
\end{eqnarray}
Two sets of moments 
\begin{eqnarray}
P^v=\langle H^v(v) \rangle_f=\langle H^v(v) \rangle_\pi \ \ \ \ \textrm{and} \ \ \ \ \ P^w=\langle H^w(w) \rangle_g=\langle H^w(w) \rangle_\pi \ ,
\label{eq:moments_def}
\end{eqnarray}
are incorporated as the input to our setting. First, we have the moments $P^v$ upon which the probability density is to be inferred, and next, the moments $P^w$ are those associated with the reference density. We assume that the input moments $P^v$ are bounded and realizable. The latter implies that there exists at least one probability density on $\mathbb{R}^3$ with moments $P^v$ . 
Note that $P^w$ is by construction realizable, since $\bar{g}$ is given. \\ \ \\
Next, we augment the introduced optimization problem with corresponding moment constraints. Let us consider the Lagrange multipliers $\lambda=[\lambda^v_1,...,\lambda^v_n,\lambda^w_1,...,\lambda^w_n]^T\in \mathbb{R}^{2n}$ which enforce the moment constraints given by $P=[P^v_1,...,P^v_n,P^w_1,...,P^w_n]^T\in \mathbb{R}^{2n}$, corresponding to the polynomials $H$.  We seek the solution of minimizing the loss functional
\begin{eqnarray}
\mathcal{L}^\lambda_{\alpha}(\pi)&=&\alpha \bigg\langle C_0|v-w|^p\bigg\rangle_\pi + \bigg\langle \log(\pi)-1\bigg\rangle_\pi-\lambda_i\bigg(\langle H_i\rangle_\pi -P_i\bigg)\ ,
\label{eq:cost_functional}
\end{eqnarray}
where, and henceforth, the summation convention is assumed for repeated indices. Since the above Lagrangian is convex (this can be readily seen by taking the variational derivatives, see e.g. \cite{pavan2011general} for more technical discussion), the solution, if it exists, is unique and it should lie at the extremum. Therefore by setting the variational derivatives to zero, we arrive at the WE distribution function  
\begin{eqnarray}
\label{eq:pi}
\pi(v,w)&=&\exp\bigg(\lambda_iH_i(v,w)-{\alpha}C_0\lvert v-w \rvert ^p\bigg) \ .
\end{eqnarray}
The above solution still depends on the unknown Lagrange multipliers. By inserting the above result for $\pi$ back in the minimization problem \eqref{eq:cost_functional}, we get the dual formulation  (see e.g. \cite{boyd2004convex}), leading to
\begin{eqnarray}
\lambda &=&\argminA_{\lambda^*\in\mathbb{R}^{2n}}\left\{\int_{\mathbb{R}^3 \times \mathbb{R}^3}\exp\bigg(\lambda^*_i H_i-\alpha C_0|v-w|^p \bigg)\  \text{dvdw} -\lambda^*_i P_i \right\}
\label{eq:opt}
\end{eqnarray}
which gives us the dual optimization problem for finding Lagrange multipliers and hence delivers the closure. However prior to that, we need to fix the model constants and references. 
\begin{enumerate}
\item {\it Reference density $\bar{g}$:} The equilibrium distribution plays a central role in the physics of gas kinetic systems. On one hand, it maximizes the entropy in the ergodic limit, and on the other hand, the molecular system tries to minimize its transportation cost with respect to the equilibrium (due to the least action principle). As the equilibrium state offers a suitable candidate for the reference density, we set 
\begin{eqnarray}
\bar{g}(w)&=&\mathcal{M}_w(\rho,u,\theta):=\frac{\rho}{(2\pi\theta)^{3/2}}\exp\left(-\frac{(w_i-u_i)(w_i-u_i)}{2\theta}\right)\ ,
\end{eqnarray}
where $\rho=\langle {\bm 1} \rangle_f$ is the density, $u=\langle v \rangle_f/\rho$ is the bulk velocity, and $\theta=1/3\langle |v-u|^2 \rangle_f/\rho$. The latter is related to the temperature $T$ via $\theta=k_bT/m$ where $m$ is the molecular mass and $k_b$ is the Boltzmann constant, while $\rho=nm$, where $n$ denotes the number density. Therefore, in the considered setup, we assume that every element of $P^w$ is generated from the Maxwellian with parameters $\rho$, $u$, and $\theta$ (the moments related to the collisional invariants)  chosen to match those of $f$. In what follows, unless mentioned otherwise, in addition to the above parameters the WE closure is augmented with stress and heat-flux information, namely
 \begin{eqnarray}
 \Sigma_{ij}&=&\theta\langle \mathcal{A}_{ij}(v) \rangle_f \ \ \ \ \ \textrm{and} \ \ \ \ \ q_i=\theta^{3/2}\langle \mathcal{B}_{i}(v) \rangle_f,
 \end{eqnarray}
 respectively, where 
 \begin{eqnarray}
\mathcal{A}_{ij}(v)&=&\frac{1}{\theta}(v_i-u_i)(v_j-u_j)-\frac{1}{3\theta}|v-u|^2\delta_{ij}\ \\
\textrm{and} \ \ \ \mathcal{B}_i(v)&=&\left(\frac{1}{2\theta}|v-u|^2-\frac{5}{2}\right)\frac{v_i-u_i}{\sqrt{\theta}} \ .
\end{eqnarray} 
\item {\it Normalization factor $C_0$:} For dimensional consistency, it is necessary to normalize the introduced distance $\langle |u-v|^p \rangle_\pi$. We adopt $\sqrt{\theta_0}$ (a reference thermal velocity) as a normalization for the velocity space. Furthermore by introducing a small parameter (empirically chosen) $\epsilon_0=10^{-3}$, we  set
\begin{eqnarray}
C_0&=&\epsilon_0 \ {{\theta_0}^{-p/2}} \ .
\end{eqnarray}
\item {\it Mixing coefficient $\alpha$:} Due to the physical observation that the equilibrium state has the maximum entropy, it is desirable to choose $\alpha$ such that $\mathcal{L}_\alpha\to \mathcal{H}$ as the system approaches equilibrium. We consider
\begin{eqnarray}
\label{eq:alpha}
\alpha&= &|\bar{P}^v-\bar{P}^w|^2/|\bar{P}^w|^2,
\end{eqnarray}
where 
\begin{eqnarray}
\bar{P}^v=\left\langle H^v\left(v\theta_0^{-1/2}\right)\right\rangle_{f} \ \ \ \ \textrm{and}
\ \ \ \ \bar{P}^w=\left\langle H^w\left(w\theta_0^{-1/2}\right)\right\rangle_{g}  
\end{eqnarray}
are moments computed with normalized velocities. 
This choice of $\alpha$ guarantees that the Wasserstein term vanishes in the vicinity of the equilibrium distribution. 
\item {\it Exponent p:} By comparing the exponent of $\pi$ to the maximum-entropy ansatz of the form $\exp(\lambda_i H_i)$, we observe the impact of the Wasserstein distance on the regularity of $\pi$. Suppose the largest power of considered polynomials is $k$, i.e. $H$ grows as $v^k$ for large $v$. Then by choosing 
\begin{eqnarray}
p&=&k+1\ ,
\end{eqnarray}
the Wasserstein term $-\alpha C_0|v-w|^p$ suppresses the exponential growth of the tails at infinity. 
\end{enumerate}
In \ref{proof:L1}, we show that the introduced WE with the adopted constants belongs to the set
\begin{eqnarray}
\label{eq:set-admit}
K_\pi&:=&\left\{\pi\ge 0, \int_{\mathbb{R}^3\times \mathbb{R}^3} \pi(v,w)\  \textrm{dvdw}<+\infty, \langle H \rangle_\pi <+\infty \right\} 
\end{eqnarray} 
and therefore the degeneracy issue faced by MED \cite{hauck2008convex} is avoided.
Furthermore, it will be shown in the follow-up section that WE recovers the Euler/Navier-Stokes-Fourier system of hydrodynamic equations in the equilibrium limit.
\section{Recovery of Hydrodynamic Models}
\label{sec:MomentSystemLimit}
\noindent In this section we investigate the consistency between the WE closure and its hydrodynamic counterpart in the equilibrium limit. Conceptually, this can be seen by noting that the contribution of the Wasserstein term vanishes in the equilibrium vicinity. As a result, the joint density $\pi$ degenerates into $\pi=fg$ where $g$ is Maxwellian and $f$ has a MED form. Hence, consistency with hydrodynamic models can be shown in the same fashion as in the case of MED \cite{levermore1996moment}. 
\\ \ \\
 Suppose we have the kinetic evolution equation of the form
\begin{eqnarray}
\label{eq:kinetic}
\partial_t \bar{f}(v|x,t)+\partial_{x_i}(v_i\bar{f}(v|x,t))&=&\mathcal{C}[\bar{f}(v|x,t)],
\end{eqnarray}
where $\mathcal{C}[\ .\ ]$ is the collision operator, e.g. the Boltzmann collision operator \cite{Cercignani}, BGK \cite{bhatnagar1954model,shakhov1968generalization,holway1966new}, or Fokker-Planck \cite{Lebowitz1960,Jenny2010,gorji2021entropic,mathiaud2016fokker}. Similarly, we have an evolution equation for the reference density 
\begin{eqnarray}
\partial_t \bar{g}(w|x,t)+\partial_{x_i}(w_i
\bar{g}(w|x,t))&=&\mathcal{C}[\bar{g}(w|x,t)],
\end{eqnarray}
where the right-hand-side becomes zero due to the choice $\bar{g}(w|x,t)=\mathcal{M}_w\left(\rho(x,t),u(x,t),\theta(x,t)\right)$.
The corresponding moment hierarchy for polynomials $H^v(v)\in\mathbb{M}^v$ reads
\begin{eqnarray}
\label{eq:ms}
\partial_t \langle H_i^v(v) \rangle_{\bar{f}} +\partial_{x_j}\langle v_jH^v_i(v) \rangle_{\bar{f}}&=&\int_{\mathbb{R}^3} H^v_i(v)\ \mathcal{C}\left[\bar{f}(v|x,t)\right] \ \textrm{dv} 
\end{eqnarray}
and
\begin{eqnarray}
\label{eq:ms_g}
\partial_t \langle H_i^w(w) \rangle_{\bar{g}} +\partial_{x_j}\langle w_jH^w_i(w) \rangle_{\bar{g}}&=&0 \ .
\end{eqnarray}
However the above system is not closed, since the polynomials underlying $\langle v_jH^v_i(v) \rangle_{\bar{f}}$ (and possibly its right-hand-side) may not belong to $\mathbb{M}^v$.  In order to proceed, let us approximate $\bar{f}$ via
\begin{eqnarray}
\label{eq:marginal_f}
f(v|x,t)&=&\int_{\mathbb{R}^3} \pi(v,w|x,t) \ \textrm{dw}
\end{eqnarray}
where $\pi$ comes from Eq.~\eqref{eq:pi} for each $x$ and $t$.
The marginal
\begin{eqnarray}
\label{eq:marginal_g}
g(v|x,t)=\int_{\mathbb{R}^3} \pi(v,w|x,t) \ \textrm{dv},
\end{eqnarray}
 is prescribed by $g=\bar g$. In other words, $g$ is taken to be the local equilibrium distribution, which provides a reasonable starting point for approximating the non-equilibrium distribution \cite{Cercignani}.
\\ \ \\
To find the hydrodynamic limit of Eq.~\eqref{eq:marginal_f} (and hence Eq.~\eqref{eq:kinetic}), we consider the expansion with respect to the small parameter $\epsilon$ (similar to the Knudsen number, $\mathrm{Kn}$)
\begin{eqnarray}
\pi&=&\pi^{(0)}+\epsilon \pi^{(1)}+\epsilon^2 \pi^{(2)}+... 
\end{eqnarray}
for the joint density which is linked to the marginals
\begin{eqnarray}
f^{(k)}&=&\int_{\mathbb{R}^3}\pi^{(k)} \textrm{dw} \ \ \ \ \ \ \textrm{and} \ \ \ \ \ \ 
g^{(k)}=\int_{\mathbb{R}^3}\pi^{(k)} \textrm{dv}.
\end{eqnarray}
In the limit $\epsilon=0$ (equilibrium) we have $\pi^{(0)}(v,w)=f^{(0)}(v)g^{(0)}(w)$ where both $f^{(0)}=\mathcal{M}_v(\rho,u,\theta)$ and $g^{(0)}=\mathcal{M}_w(\rho,u,\theta)$ are Maxwellians with the same moments. Given the choice $g=\bar g=\mathcal{M}_w(\rho,u,\theta)$, this implies 
$g^{(i)}=0$  and $\int_{\mathbb{R}^3} f^{(i)}\textrm{dv} =0$ for $i\ge 1$.
Similar expansions hold for the Lagrange multipliers 
\begin{eqnarray}
\lambda_i&=&\lambda_i^{(0)}+\epsilon \lambda_i^{(1)}+\epsilon^2\lambda_i^{(2)}+... 
\end{eqnarray}
and moments
\begin{eqnarray}
P^v&=&{P^v}^{(0)}+\epsilon {P^v}^{(1)}+\epsilon^2 {P^v}^{(2)}+... \ .
\end{eqnarray}
By virtue of ${P^v}^{(0)}=P^w$, and assuming $\theta_0=1$ for notational simplicity and without loss of generality, Eq.~\eqref{eq:pi} reduces to
\begin{eqnarray}
    \pi &=& \exp\left(  H_i \lambda_i ^{(0)}+  \epsilon H_i \lambda_i ^{(1)}+... +  C_0\frac{|\cancel{P^{v(0)}}+\epsilon P^{v(1)}+...-\cancel{P^{w}}|^2}{|P^{w}|^2} |v-w|^p \right)
    \\
  &=& \exp{\left( H_i \lambda_i^{(0)} \right)} 
 \overbrace{\exp \left( \epsilon H_i \lambda_i^{(1)} + \epsilon^2 C_0 \frac{|P^{v(1)}|^2}{|P^w|^2 } |v-w|^p + ... \right)}^{{\delta}}
 \\
  &=& \pi^{(0)}\left( 1 + \epsilon H_i \lambda_i^{(1)} + \mathcal{O}(\epsilon^2) \right) \ \ \ \textrm{(using Taylor expansion of $\delta$)}
\end{eqnarray}
which provides us with the first-order approximation
\begin{eqnarray}
\pi^{(1)}&=& \pi^{(0)}\left( H_i\lambda^{(1)}_i \right)~.
\label{eq:pi^1}
\end{eqnarray}
We note that the Wasserstein term does not appear in the zeroth- and first-order approximations. By taking the marginal of $\pi^{(1)}$ we obtain
\begin{eqnarray}
f^{(1)}&=&  \int_{\mathbb{R}^3}\pi^{(0)} H_i\lambda^{(1)}_i  \textrm{dw}=  f^{(0)} \overbrace{\left(H_i^v{\lambda_i^v}^{(1)}+\langle H_i^w\rangle_{g^{(0)}}{\lambda_i^w}^{(1)}\right)}^{\delta f^{(1)}} \ \ \ \ \ \textrm{and}
\label{eq:f1}\\
 g^{(1)}&=&  \int_{\mathbb{R}^3}\pi^{(0)} H_i\lambda^{(1)}_i  \textrm{dv}=  g^{(0)}\overbrace{\left(H_i^w{\lambda_i^w}^{(1)}+\langle H_i^v\rangle_{f^{(0)}}{\lambda_i^v}^{(1)}\right)}^{\delta g^{(1)}}.
\end{eqnarray}
 However, since $\int_{\mathbb{R}^3} f^{(1)}\textrm{dv} =0$ we obtain
\begin{eqnarray}
\langle H_i^v \rangle_{f^{(0)}}{\lambda_i^v}^{(1)}&=&-\langle H_i^w \rangle_{g^{(0)}}{\lambda_i^w}^{(1)}
\end{eqnarray}
and thus
\begin{eqnarray}
\delta g^{(1)}&=&(H_i^w-\langle H_i^w \rangle{_{g^{(0)}}}){\lambda_i^w}^{(1)}\ .
\end{eqnarray}
The latter expression implies that ${\lambda^w}^{(1)}=0$, while from the former we obtain 
\begin{eqnarray}
\delta f^{(1)}=H_i^v{\lambda_i^v}^{(1)} \label{eq:delta_f1} \ .
\label{eq:df1_Hlam}
\end{eqnarray}
Next we need to show that there exists Lagrange multipliers for which the moment hierarchy corresponding to the kinetic equation \eqref{eq:kinetic} 
converges to the Euler system for $\bar{f}=f^{(0)}$ and to the NSF system for $\bar{f}=f^{(0)}+\epsilon f^{(1)}$. The former is trivial to check since $f^{(0)}=\mathcal{M}_v(\rho,u,\theta)$; the details for the latter are provided in \ref{proof:NSF}.  
\section{Stochastic Representation}

 \label{sec:StochRep}
 
\noindent  The optimization problem given by Eq.~\eqref{eq:opt} can be solved using nonlinear solvers such as the Newton-Raphson method \cite{schaerer201735}. Unfortunately, this approach becomes prohibitive in high-dimensional settings. While machine learning methodologies have been pursued recently \cite{sadr2020gaussian,schotthofer2022structure}, efficient and reliable numerical schemes for the affordable computation of Lagrange multipliers for practical scenarios have yet to be developed. Here, we present a new approach based on the Fokker-Planck interpretation of our introduced closure solution. Namely, we reset the problem within a stochastic framework where WE is the stationary solution of the associated Stochastic-Differential-Equations (SDEs). In what follows, we propose this stochastic representation and relegate the more technical details to \ref{proof:MC}. \\ \ \\
Recall that $\Omega$ is the physical space under consideration. Furthermore, let us consider a probability space $(\mathcal{X},\mathbb{P})$ with the sample space $\mathcal{X}$ and the law $\mathbb{P}$. Suppose $Z^{t,x}=(V^{t,x}_1,V^{t,x}_2,V^{t,x}_3,W^{t,x},W^{t,x}_2,W^{t,x}_3)^T$, with $Z^{t,x}: \mathcal{X}\to \mathbb{R}^6$, is a random variable, indexed by the time $t\in\mathbb{R}^{+}$ and the position $x\in\Omega$.  We focus on an evolution, governed by the following It\^{o} SDEs
 \begin{eqnarray}
 \text{d}V_i^{t,x}&=&-\tilde{\lambda}^v_j(x,t)\partial_{v_i}  H^v_j(V^{t,x})\text{d}t-C_0\tilde{\alpha}(x,t) p\left(V^{t,x}_i-W^{t,x}_i\right)|V^{t,x}-W^{t,x}|^{p-2}\text{d}t+\sqrt{2}\ \text{d}B^{v,t}_i \nonumber \\
 \textrm{and} \ \ \ \ \ 
 \text{d}W_i^{t,x}&=&-\tilde{\lambda}^w_j(x,t)\partial_{w_i} H^w_j(W^{t,x})\text{d}t-C_0\tilde{\alpha}(x,t) p\left(W^{t,x}_i-V^{t,x}_i\right)|V^{t,x}-W^{t,x}|^{p-2}\text{d}t+\sqrt{2}\ \text{d}B^{w,t}_i, \nonumber \\
 \label{eq:SDEs}
 \end{eqnarray}
 where $\text{d}B^t=[\text{d}B^{v,t}_1,\text{d}B^{v,t}_2,\text{d}B^{v,t}_3,\text{d}B^{w,t}_1,\text{d}B^{w,t}_2,\text{d}B^{w,t}_3]^T$ is a six-dimensional Brownian (Wiener) process in time. Besides,  $\tilde{\lambda}=[\tilde{\lambda}_1^v,...,\tilde{\lambda}_n^v,\tilde{\lambda}_1^w,...,\tilde{\lambda}_n^w]^T$ and $\tilde{\alpha}$ are the Lagrange multipliers and normalization factor (see Eq.~\eqref{eq:alpha}), respectively, which both correspond to a moment vector $\tilde{P}=[\tilde{P}^v_1,...,\tilde{P}^v_n,\tilde{P}^w_1,...,\tilde{P}^w_n]^T$.  
 It is straightforward to show that the density corresponding to the law of $Z$ converges to the WE density $\pi$, as $t\to\infty$ (see \ref{proof:MC}). \\ \ \\
 In order to set up the numerical scheme, we also need to devise an algorithm to find the optimal Lagrange multipliers. 
 As will be shown below, by coupling the moments to the Lagrange multipliers, we can construct a time marching scheme to update the Lagrange multipliers towards their optimal values. \\ \ \\
 A time marching scheme where the coupling between estimated Lagrange multipliers $\tilde{\lambda}$ and the moments 
 \begin{eqnarray}
 \tilde{P}^v_l(x,t)&=& \mathbb{E}[H^v_l(V^{t,x})] \ \ \ \ \ \ \ \textrm{and}
  \ \ \ \ \ \ \
\tilde{P}^w_l(x,t)= \mathbb{E}[H^w_l(W^{t,x})]
 \end{eqnarray}
 is exploited can be built as follows (here and henceforth $E[A]$ is the expectation of $A$ with respect to the law $\mathbb{P}$, and not the parameters $x$ and $t$). From SDEs \eqref{eq:SDEs}, we obtain
\begin{eqnarray}
\label{eq:consist}
\partial_t \tilde{P}^{v}_l(x,t)&=&-\mathbb{E}\bigg[\partial_{v_i}\bigg(H^v_k(V^{t,x}) \tilde{\lambda}^v_k(x,t)-\tilde{\alpha}(x,t)C_0|V^{t,x}-W^{t,x}|^{p}\bigg)\partial_{v_i}H_l^v(V^{t,x})
\nonumber \\ &&\ \ \ \ \ \ \ -\partial^2_{v_iv_i} H_l^v(V^{t,x}) \bigg] \nonumber \\
\textrm{and} \ \ \ \ \ \ \partial_t \tilde{P}^{w}_l(x,t)&=&-\mathbb{E}\bigg[\partial_{w_i}\bigg(H^w_k(W^{t,x}) \tilde{\lambda}^w_k(x,t)-\tilde{\alpha}(x,t)C_0|V^{t,x}-W^{t,x}|^{p}\bigg)\partial_{w_i}H_l^w(W^{t,x}) \nonumber \\
&&\ \ \ \ \ \ \ -\partial^2_{w_iw_i} H_l^w(W^{t,x}) \bigg].
\end{eqnarray}
The above equations connect the updates in $\tilde{P}$ to $\tilde{\lambda}$. By requiring that the moments approach the target values according to the linear relaxation law
 \begin{eqnarray}
 \label{eq:rel}
 {\partial_t}\tilde{P}&=&\frac{1}{\tau}\left(P-\tilde{P}\right),
 \end{eqnarray}
 where $\tau>0$ is an input parameter that controls the convergence speed,  Eqs.~\eqref{eq:consist} can be converted into a linear system for $\tilde{\lambda}$. To this end, we define
 \begin{eqnarray}
 \mathscr{R}(x,t)&=&\begin{bmatrix}
 \mathscr{R}^v(x,t) \\
 \mathscr{R}^w(x,t)
 \end{bmatrix} 
 \end{eqnarray}
 where
 \begin{eqnarray}
 \mathscr{R}^v_i&=& \tau^{-1}\left(P_i^v-\tilde{P}_i^v\right)-\mathbb{E}\left[\partial^2_{v_jv_j}H_i^v(V)+pC_0\tilde{\alpha}\partial_{v_j}H_i^v(V)(V_j-W_j)|V-W|^{p-2}\right]  \nonumber
 \end{eqnarray}
 and
 \begin{eqnarray}
 \mathscr{R}^w_i&=&\tau^{-1}\left(P_i^w-\tilde{P}_i^w\right)-\mathbb{E}\left[\partial^2_{w_jw_j}H_i^w(W)+pC_0\tilde{\alpha}\partial_{w_j}H_i^w(W)(W_j-V_j)|V-W|^{p-2}\right] \nonumber \ ,
 \end{eqnarray}
 where for brevity we omit the $x$ and $t$ dependency in the notation.
 Next, we define 
 \begin{eqnarray}
 \mathscr{A}(x,t)&=& \begin{bmatrix}
 \mathscr{A}^v(x,t) \\
 \mathscr{A}^w(x,t)
 \end{bmatrix}
 \label{eq:matrix_to_invert}
 \end{eqnarray}
 where
 \begin{eqnarray}
 \mathscr{A}^v_{ij}&=& \mathbb{E}\left[\partial_{v_k}{H_i^v(V)}\partial_{v_k}{H_j^v(V)}\right] \  \ \ \ \ \ \ \textrm{and} \  \ \ \ \ \ \
 \mathscr{A}^w_{ij}= \mathbb{E}\left[\partial_{w_k}{H_i^w(W)}\partial_{w_k}{H_j^w(W)}\right] \nonumber
 \end{eqnarray}
 which leads to
 \begin{eqnarray}
 \label{eq:ls}
 \tilde{\lambda}(x,t)&=&\mathscr{A}^{-1}(x,t)\mathscr{R}(x,t).
 \label{eq:WE_system_Lagrange_mult}
 \end{eqnarray}
 Therefore by simulating the devised SDEs \eqref{eq:SDEs} and solving the linear system \eqref{eq:ls} we get updated values for $\tilde{\lambda}$. More details, including consistency arguments, can be found in \ref{proof:MC}. 

\section{Numerical Algorithm and Validation}

\label{sec:SolAlg_SanityCheck}



\subsection{Solution algorithm and validation}
\label{sec:sol_alg}
In this section, we propose a stochastic algorithm for creating samples of the WE closure given moments of the unknown distribution $P^v$. Given a convergence threshold $ \varepsilon$ and samples ($W$) of the prior distribution, Algorithm~\ref{alg:WE} provides samples ($V$) of the target closure problem. This algorithm requires an initial guess for $V$. In what follows, in the interest of simplicity, for this initial guess we sample $V$ from the same distribution as $W$. 
\\ \ \\
\begin{algorithm}[H]
 \caption{Algorithm for generating normalized samples of $f(v)$ from initial samples of $V$ and $W$ as well as the target moments $P^v = \mathbb{E}[H(V)]$ and  $P^w = \mathbb{E}[H(W)]$. In this work we set  $\varepsilon=10^{-3}$ and $a=10$.}
\SetAlgoLined
 -Normalize $V$ and $W$\;
 -Set $\Delta t=10^{-3}$ and $\tau = a \Delta t$\;
 \While{$||\tilde{P}^v-P^v||_2/||P^v||_2 + ||\tilde{P}^w-P^w||_2/||P^w||_2>\varepsilon$}{
  -Solve eq.~\eqref{eq:WE_system_Lagrange_mult}\;
  -Move $V$ and $W$ using Euler–Maruyama solution to eqs.~\eqref{eq:SDEs}\;  
 }
 \label{alg:WE}
\end{algorithm}

\subsection{One-dimensional moment problems and realizability}
\begin{figure}
    \centering
    \begin{tabular}{cccc}
    \includegraphics[scale=0.46]{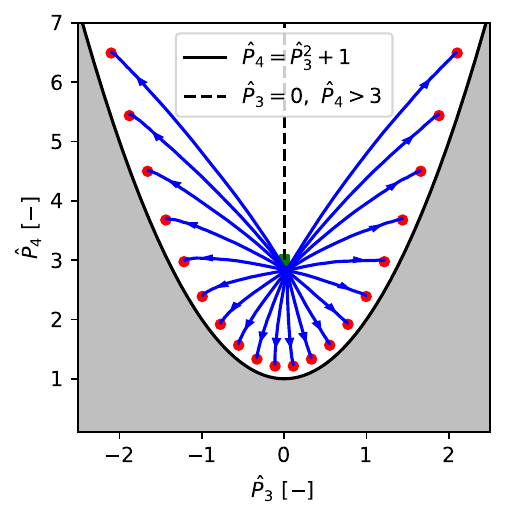} &
    \includegraphics[scale=0.46]{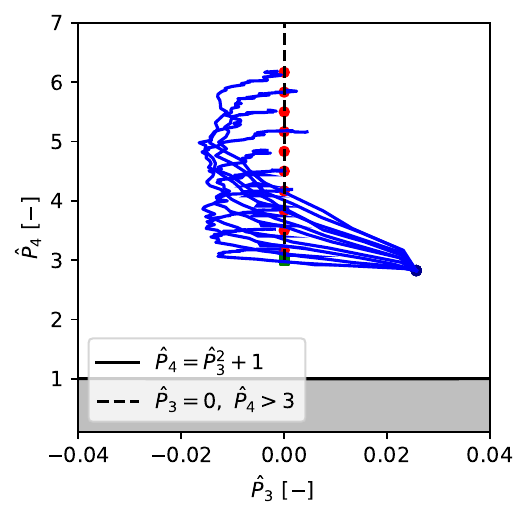} &
    \includegraphics[scale=0.46]{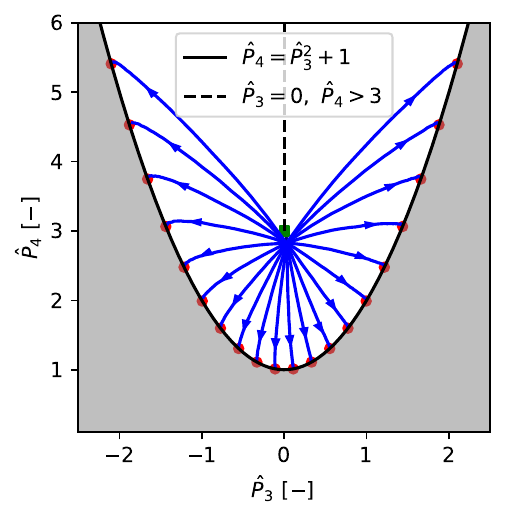} &
    \includegraphics[scale=0.46]{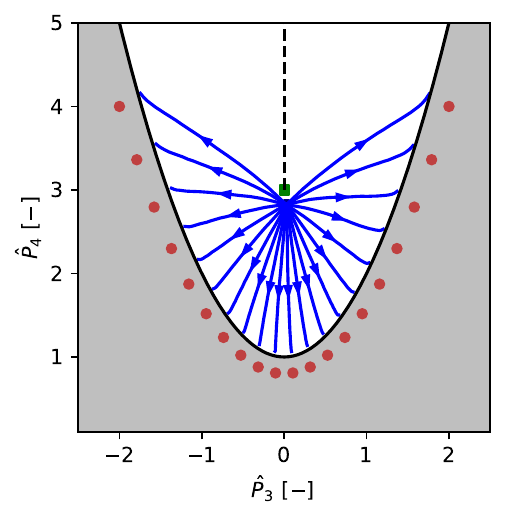}
    \\
    \includegraphics[scale=0.46]{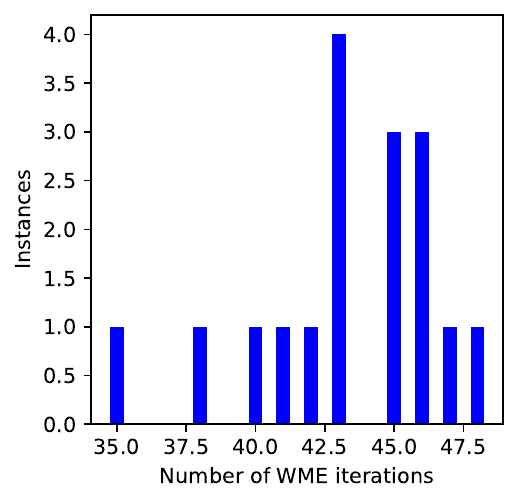}&
    \includegraphics[scale=0.46]{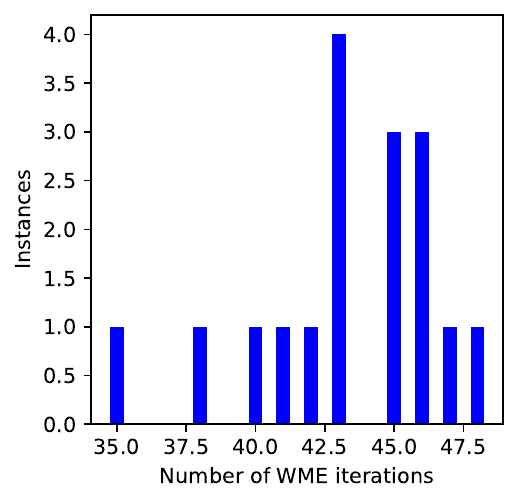}
    &
    \includegraphics[scale=0.46]{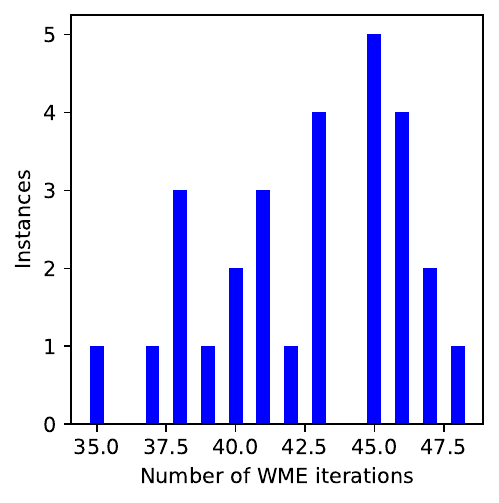}
    &
    \includegraphics[scale=0.46]{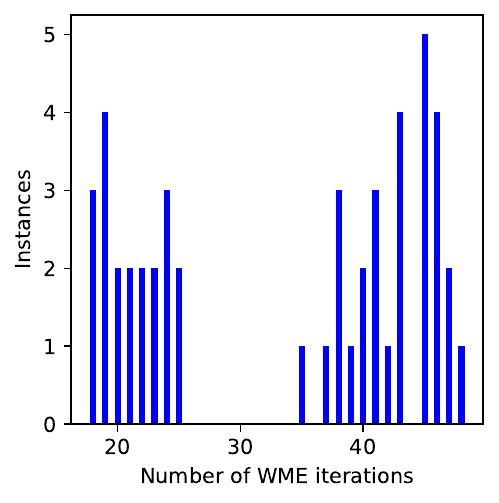}
    \\
    (a) Realizable & (b) Junk-line & (c) Limit of realizability & (d) Non-realizable
    \end{tabular}
    \caption{Convergence path in $(\hat P_3,\hat P_4)$ plane for WE solution algorithm starting from samples of the normal distribution (blue circle) and approaching the target moments (red circles) in (a) realizable region ($\hat P_4>\hat P_3^2+1$), (b) Junk line ($\hat P_3=0$ and $\hat P_4>3$) denoted by dashed lines, (c) limit of realizability ($\hat P_4=\hat P_3^2+1$) and (d) non-realizable region ($\hat P_4<\hat P_3^2+1$) shown as shaded (right). Note that, due to noise, the initial condition is slightly different from the exact location corresponding to the normal distribution (0,3), shown as a green square. The number of iterations taken for convergence, defined as reaching a relative error  of $\varepsilon=10^{-2}$, is also shown for each case.}
    \label{fig:path_4mom_num_iter}
\end{figure}

\begin{figure}
    \centering
    \begin{tabular}{cccc}
    \begin{turn}{90} \hspace{1.5cm} Realizable  \end{turn} 
    &
    \includegraphics[scale=0.5]{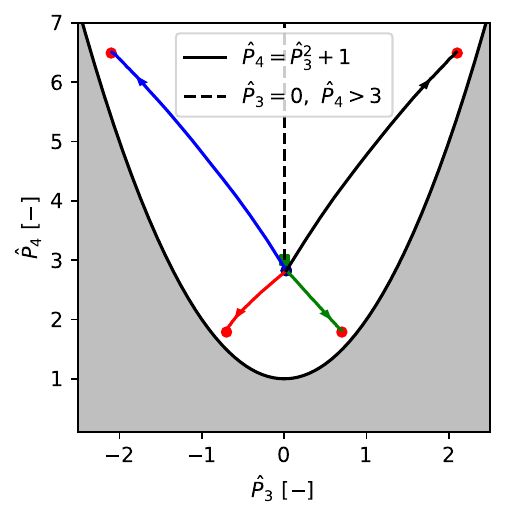} &
    \includegraphics[scale=0.5]{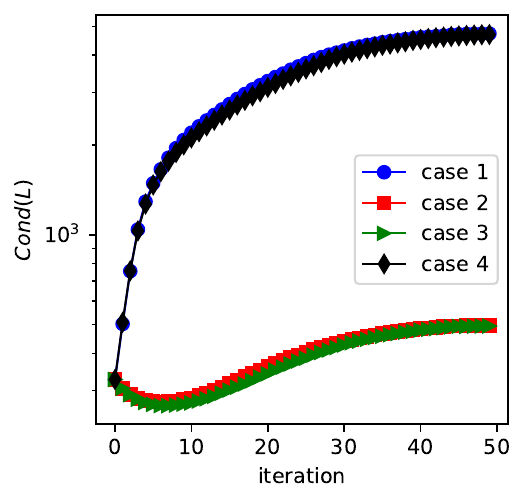} &
    \includegraphics[scale=0.5]{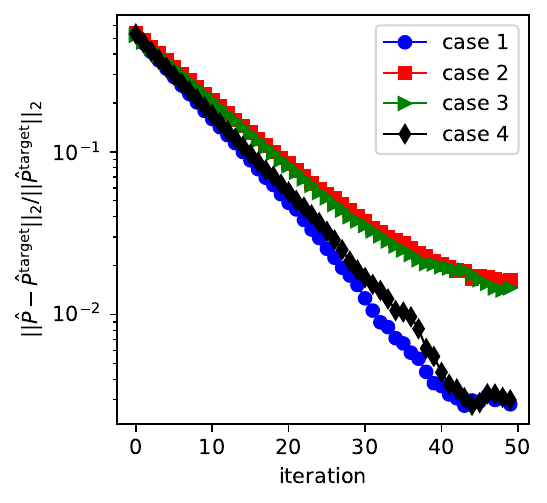}
    \\
    \begin{turn}{90} \hspace{1.5cm} Junk-line  \end{turn}  
    &\includegraphics[scale=0.5]{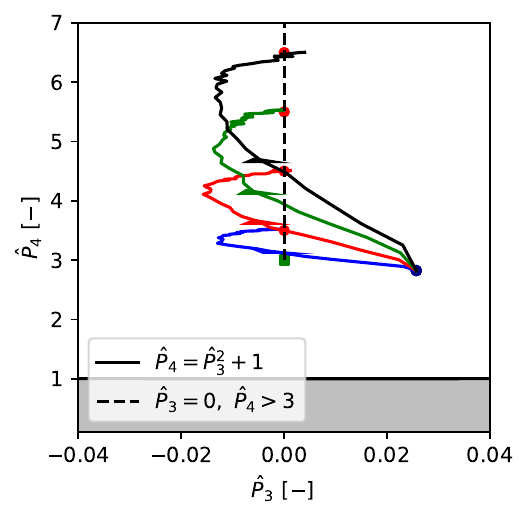} &
    \includegraphics[scale=0.5]{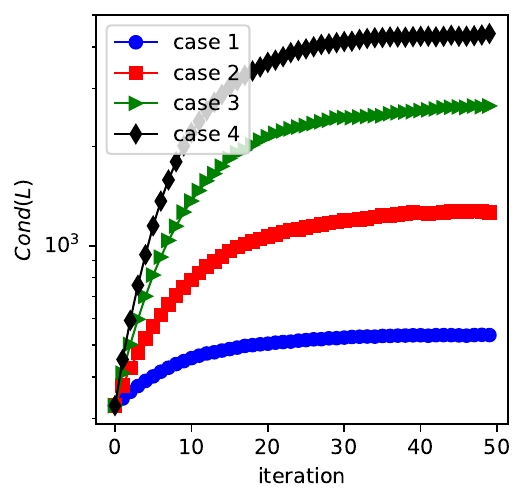} &
    \includegraphics[scale=0.5]{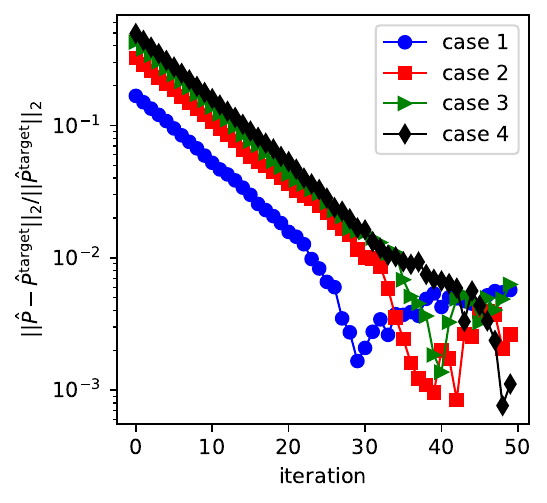}
    \\
    \begin{turn}{90} \hspace{1cm} Limit of realizability  \end{turn}  
    &\includegraphics[scale=0.5]{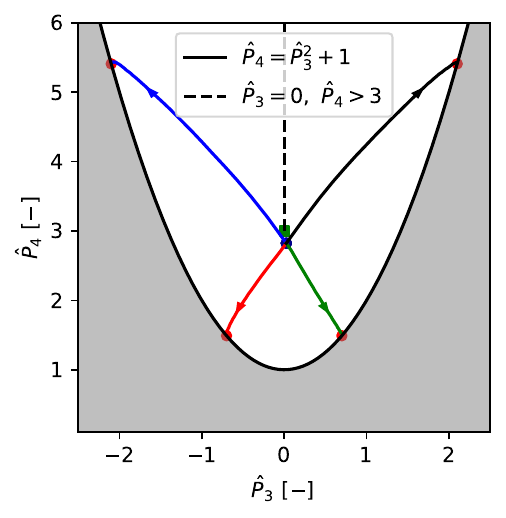} &
    \includegraphics[scale=0.5]{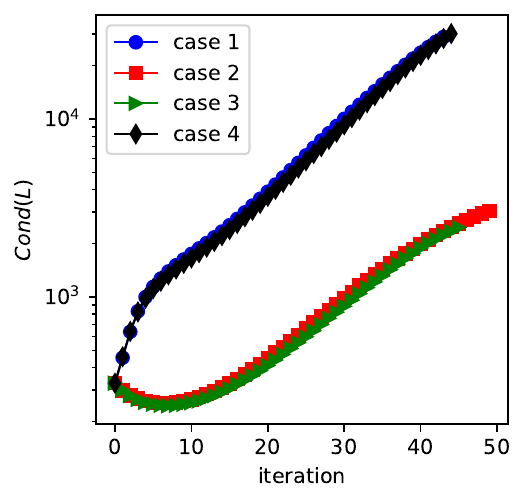} &
    \includegraphics[scale=0.5]{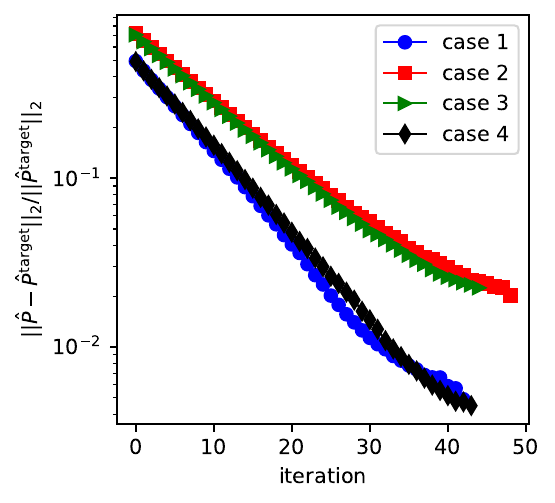}
    \\
    \begin{turn}{90} \hspace{1cm} Non-realizable  \end{turn} 
    &\includegraphics[scale=0.5]{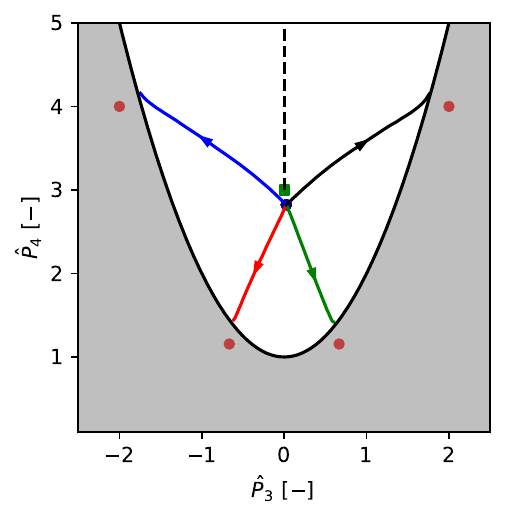} &
    \includegraphics[scale=0.5]{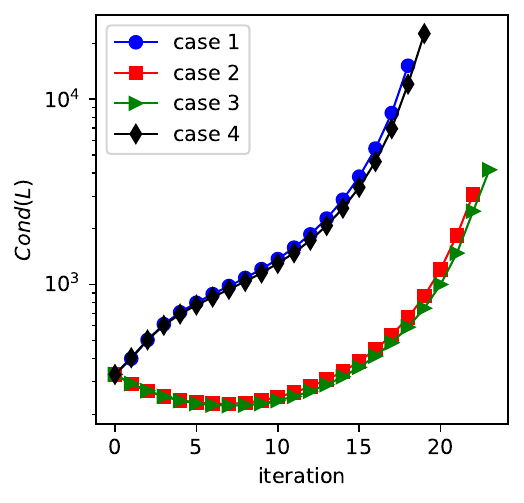} &
    \includegraphics[scale=0.5]{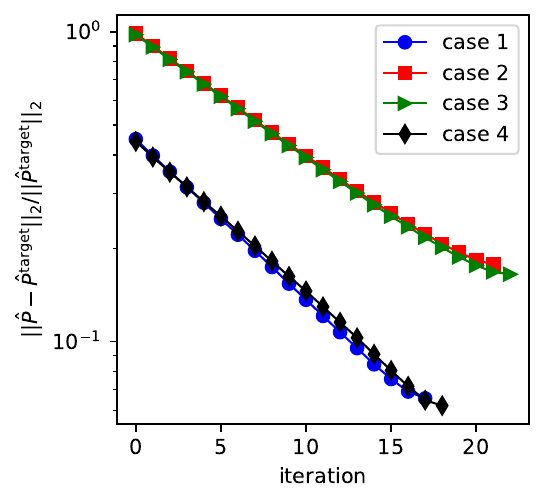}
    \end{tabular}
    \caption{WE convergence path in $(\hat P_3, \hat  P_4)$ plane (left), the evolution of condition number (middle) and relative error in moments (right) for four target points in each of the four sub-spaces defined in the text and the caption of Figure \ref{fig:path_4mom_num_iter}.}
    \label{fig:4mom_real_nonreal_junk_analysis}
\end{figure}

\begin{figure}
    \centering
    \begin{tabular}{cccc}
    \begin{turn}{90} \hspace{1cm} Realizable  \end{turn} 
    &
    \includegraphics[scale=0.5]{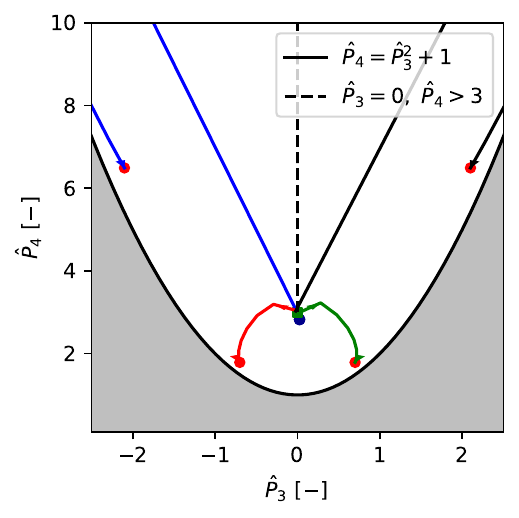} &
    \includegraphics[scale=0.5]{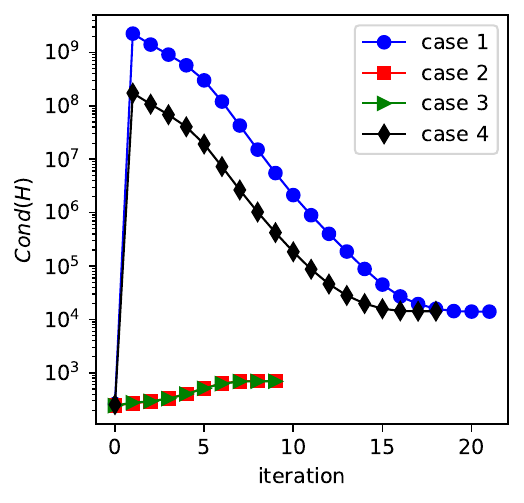} &
    \includegraphics[scale=0.5]{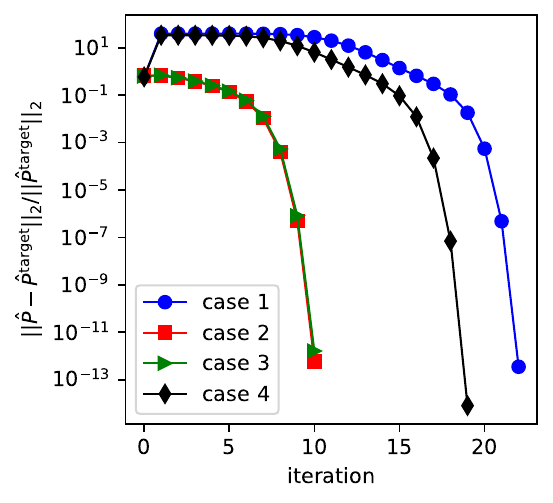}
    \\
    \begin{turn}{90} \hspace{1cm} Junk-line  \end{turn}  
    &\includegraphics[scale=0.5]{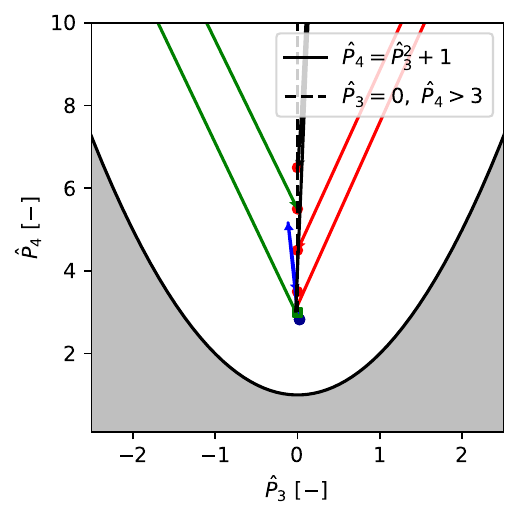} &
    \includegraphics[scale=0.5]{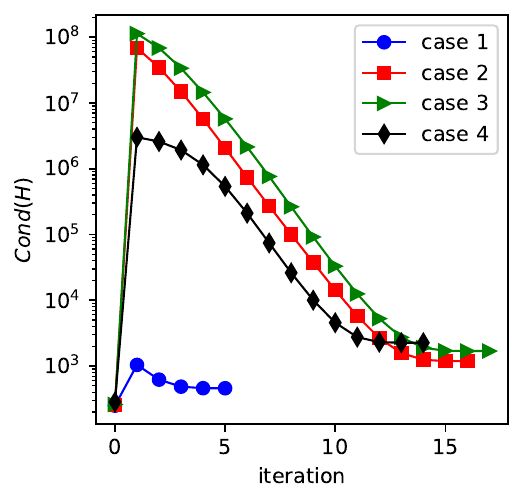} &
    \includegraphics[scale=0.5]{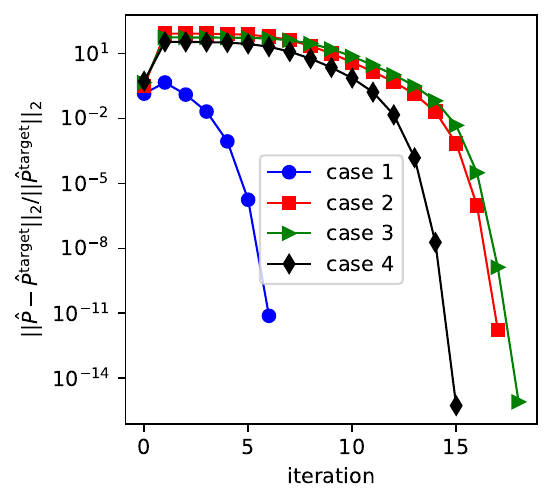}
    \\
    \begin{turn}{90} \hspace{1cm} Limit of realizablity  \end{turn}  
    &\includegraphics[scale=0.5]{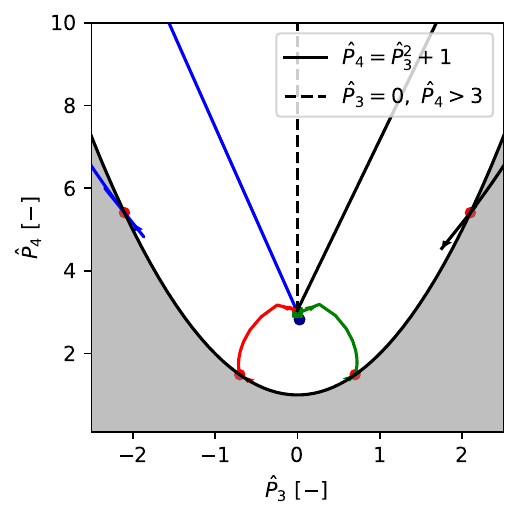} &
    \includegraphics[scale=0.5]{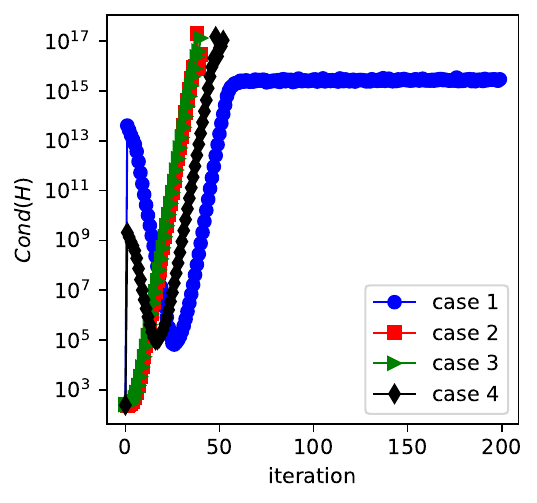} &
    \includegraphics[scale=0.5]{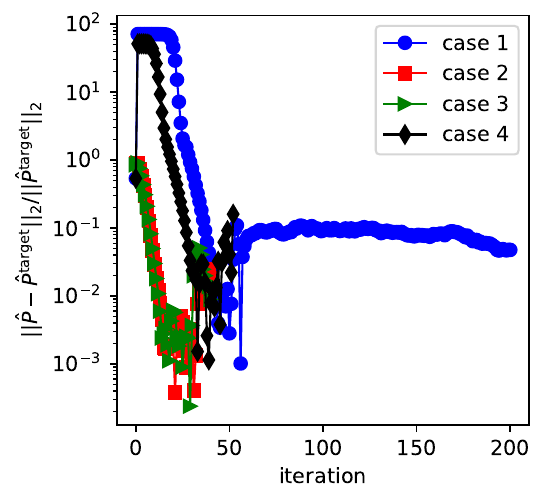}
    \\
    \begin{turn}{90} \hspace{1cm} Non-realizable  \end{turn} 
    &\includegraphics[scale=0.5]{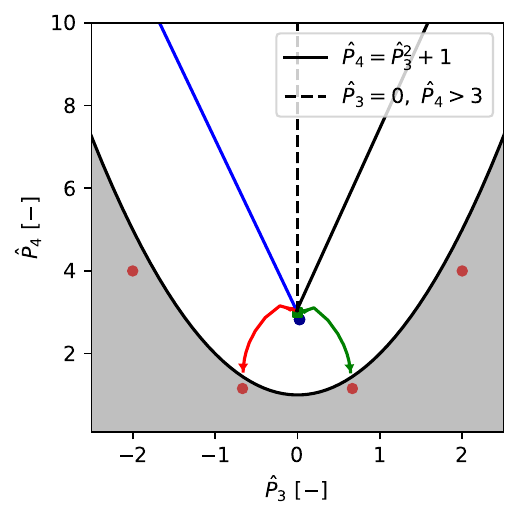} &
    \includegraphics[scale=0.5]{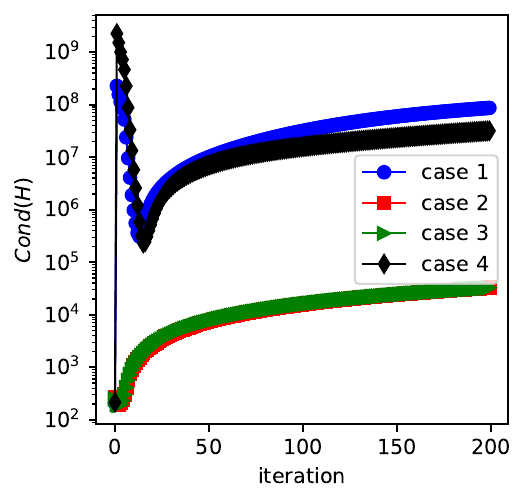} &
    \includegraphics[scale=0.5]{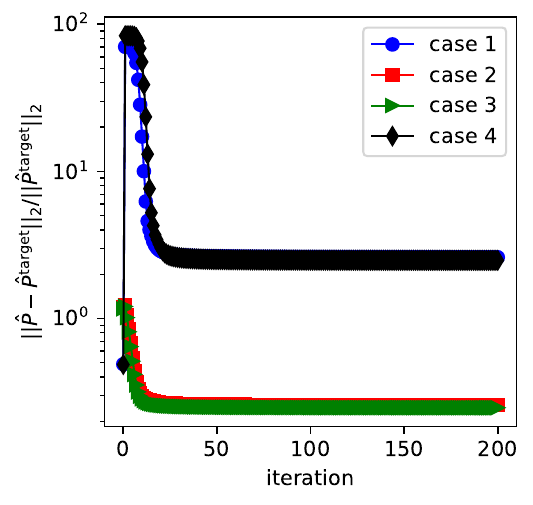}
    \end{tabular}
    \caption{MED convergence path in $(\hat P_3, \hat P_4)$ plane (left), condition number (middle) and relative error in moments (right) for the evolution of four target points in each of the four sub-spaces defined in the text and the caption of Figure \ref{fig:path_4mom_num_iter}.}
    \label{fig:MED_4mom_real_nonreal_junk}
\end{figure}

In what follows, we investigate the accuracy and efficiency of the WE closure model in one-dimensional space for a wide range of moment problems. In particular, we compare the performance of the proposed WE closure against MED solution approaches for moment problems where the latter either becomes computationally expensive or is unable to converge. 
\\ \ \\
For this purpose, we consider the moment system corresponding to the polynomial basis functions $H=\{v,v^2,...,v^{N_m}
\}$ with $N_m=4$.  It can be shown \cite{mcdonald2013affordable,akhiezer1965classical} that this moment problem is physically realizable only if 
\begin{flalign}
\hat P_4 \geq \hat P_3^2+1~
\end{flalign}
for the normalized moments $\hat P\in\mathbb{R}^4$, i.e. $\hat P = \tilde P / \rho$. The locus of points $\hat P_3=0$, $\hat P_4>3$, known as the Junk line \cite{junk1998domain}, is of particular interest here, because it cannot be reached by MED. We investigate the behavior of the methods under comparison by tracking their approach to target locations in the $\hat P_3-\hat P_4$ diagram, following initialization from a Normal distribution (point (0,3) in the same diagram). To aid the discussion, we decompose the moment space into four subspaces, namely, realizable ($\hat P_4 > \hat P_3^2+1$), Junk-line ($\hat P_3=0$ and $\hat P_4>3$), on the limit of realizability ($\hat P_4 = \hat P_3^2+1$), and physically non-realizable moments ($\hat P_4 < \hat P_3^2+1$).
\\ \ \\
Figure~\ref{fig:path_4mom_num_iter} illustrates the ability of the WE algorithm to monotonically converge to the target locations, including the Junk line and the limit of realizability. In the case of unrealizable targets, the method converges to a nearby point on the limit of realizability. 
\\ \ \\
Figure~\ref{fig:4mom_real_nonreal_junk_analysis} reports the evolution of relative error and condition number for four representative target locations from each subspace. Even in the case of non-realizable moments, we note that the proposed WE particle method can converge to a nearby solution in the realizable subspace of moments. This is achieved by stopping the process as soon as the error in moments start to increase. In the case of the Junk-line, we observe significantly more noise in the trajectory. We believe this noise is a consequence of the interplay between the destabilizing MED term and the regularizing WE contribution near the target. 
\\ \ \\
For comparison,  we also solve these moment problems using the MED with the standard dual formulation. In this approach, the gradient of the optimization problem is based on the difference between target moments and the computed moments of the current MED iterate. For simplicity, we use the normal distribution as the prior and compute the correction using the maximum cross-entropy method \cite{debrabant2017micro}.  The equivalency of the MED solution and the one obtained from maximum cross-entropy formulation follows from the uniqueness of MED for realizable distributions.  As shown in Fig.~\ref{fig:MED_4mom_real_nonreal_junk}, although MED performs well for realizable moments as well as the Junk-line, we note its failure in finding a solution on the limit of realizability and finding a realizable solution nearby a nonrealizable target. We note that the condition number of the Hessian in the MED optimization problem can be $10$ orders of magnitude larger than that of the WE method close to the limit of realizability. It is clear that WE closure provides a reliable solution at a reasonable error for target moment problems that MED suffers numerically.
\\ \ \\
Note that even though MED does not exist on the Junk line, the deployed MED algorithm can still find an estimate in its neighborhood, thanks to underlying numerical errors. In particular, we defer the divergence near the Junk line by stopping the optimization process at a tolerance of $10^{-10}$.
\subsection{Bi-modal distribution}
\begin{figure}
    \centering
    \begin{tabular}{cccc}
       \begin{turn}{90} \hspace{1cm} MED, $N_m=4,\ 6$  \end{turn} 
    &
    \includegraphics[scale=0.5]{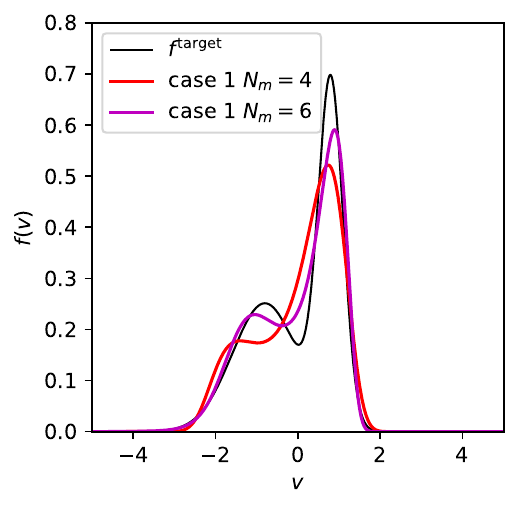} &
    \includegraphics[scale=0.5]{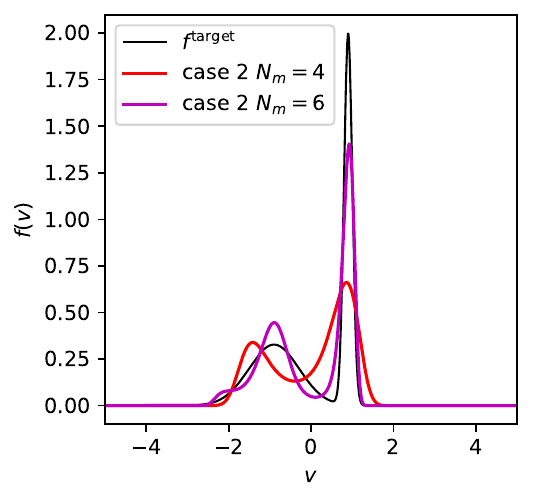} &
    \includegraphics[scale=0.5]{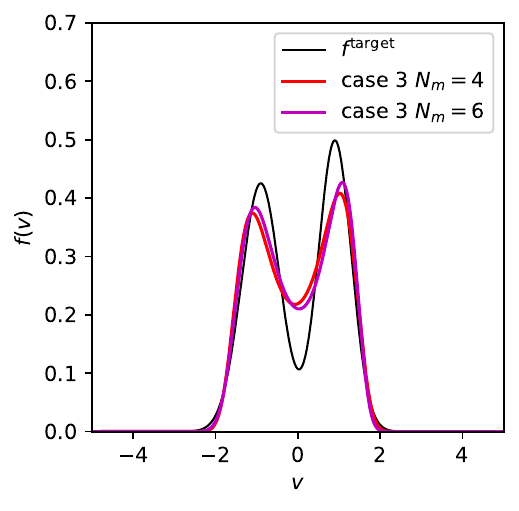}
    \\
      \begin{turn}{90} \hspace{1cm} WE, $N_m=3$  \end{turn} 
    &
    \includegraphics[scale=0.5]{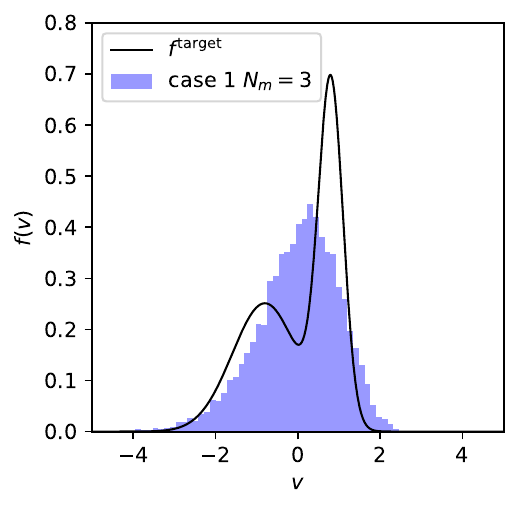} &
    \includegraphics[scale=0.5]{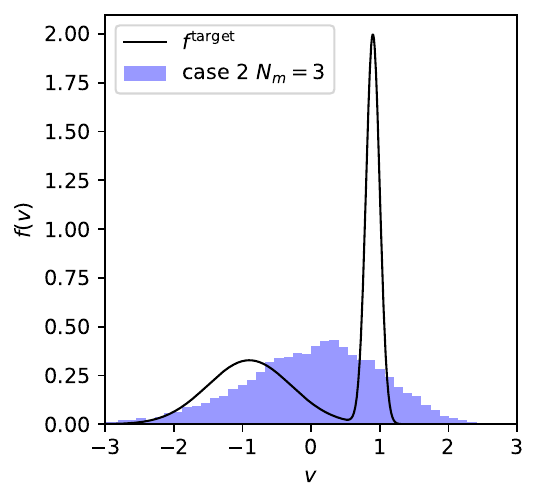} &
\includegraphics[scale=0.5]{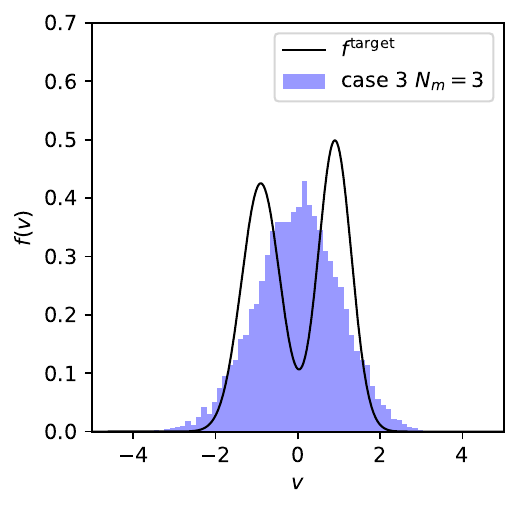}
\\
      \begin{turn}{90} \hspace{1cm} WE, $N_m=4$  \end{turn} 
    &
\includegraphics[scale=0.5]{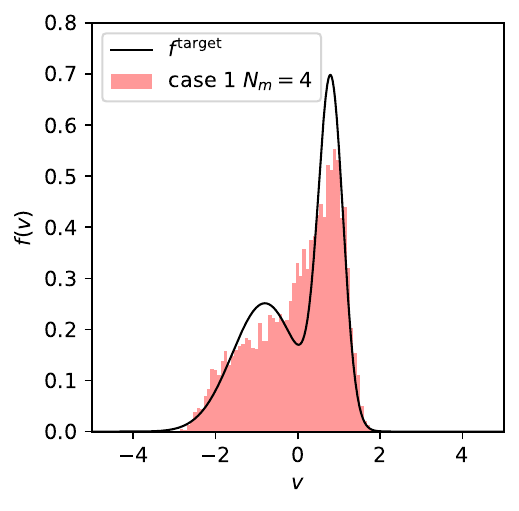} &
    \includegraphics[scale=0.5]{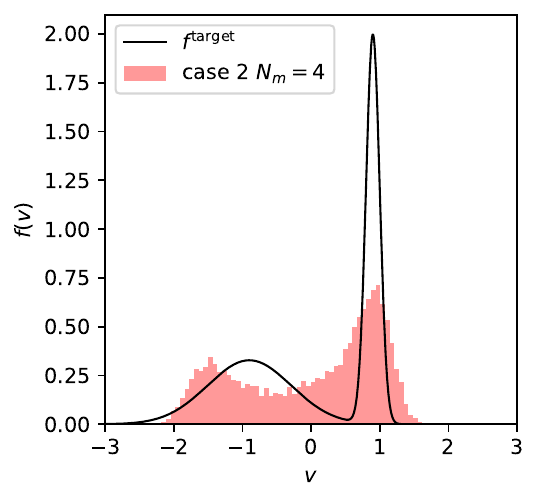} &
\includegraphics[scale=0.5]{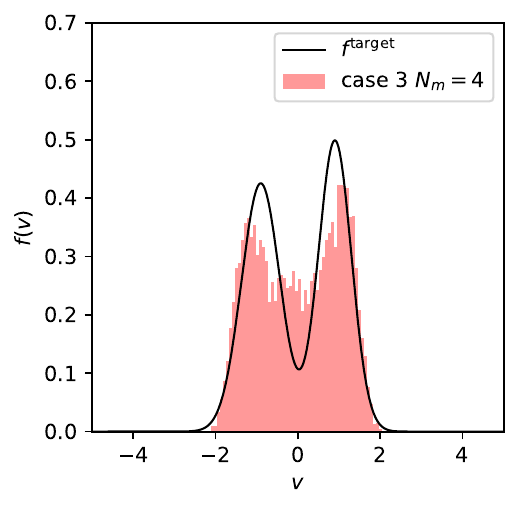}
\\
      \begin{turn}{90} \hspace{1cm} WE, $N_m=5$  \end{turn} 
    &
\includegraphics[scale=0.5]{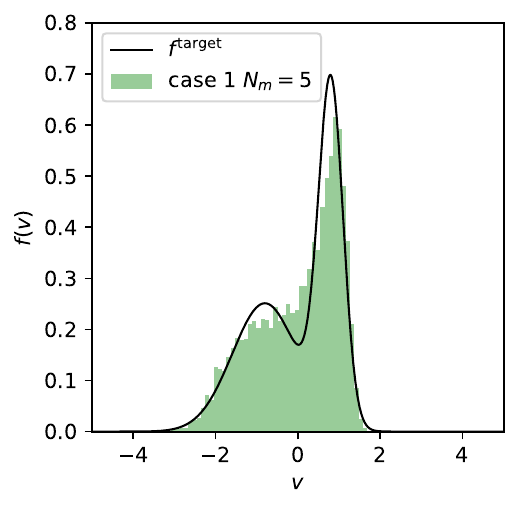} &
    \includegraphics[scale=0.5]{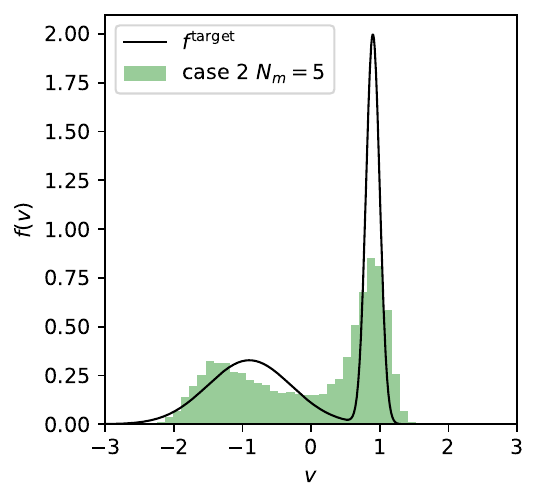} &
\includegraphics[scale=0.5]{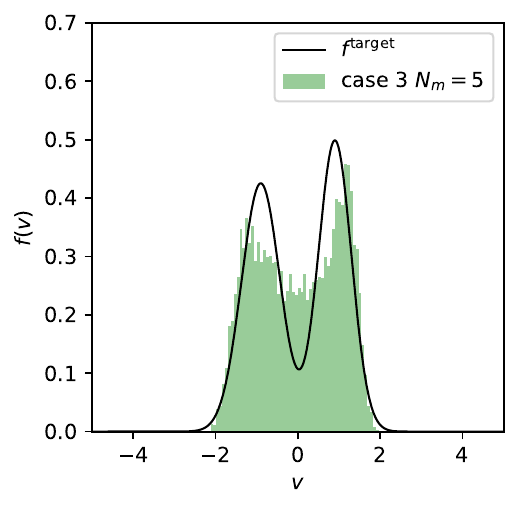}
\\
      \begin{turn}{90} \hspace{1cm} WE, $N_m=6$  \end{turn} 
    &
\includegraphics[scale=0.5]{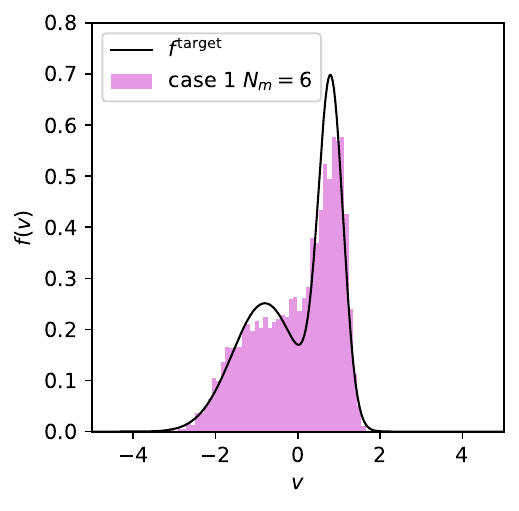} &
    \includegraphics[scale=0.5]{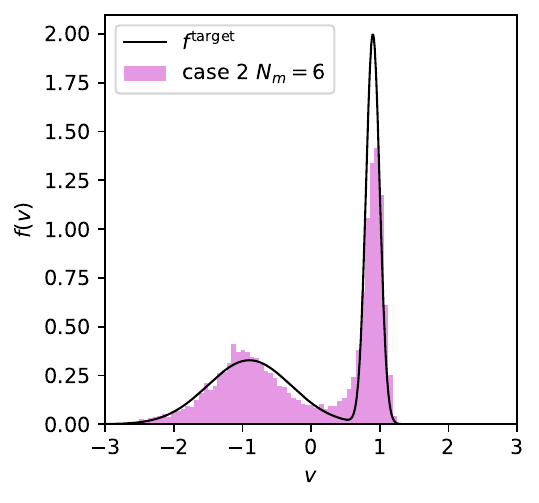} &
\includegraphics[scale=0.46]{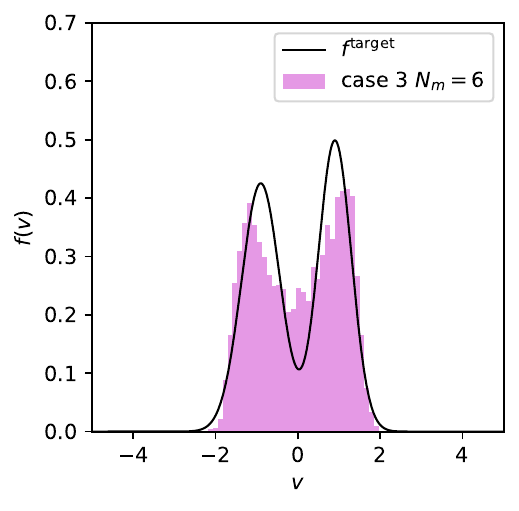}
\\
& Case 1 &  Case 2 &  Case 3
       \end{tabular}
    \caption{Approximating bi-modal distribution function (black) for cases 1-3 (see eq.~\eqref{eq:bi-modal_dist} and Table~\ref{tab:param_bi-modal-dist}) using MED with 4 and 6 moments and WE closure model that matches 3 (blue), 4 (red), 5 (green) and 6 (magma) moments.}
\label{fig:bi-mod-distr}
\end{figure}

\begin{figure}
    \centering
    \hspace{-1.38cm}
    \begin{tabular}{ccc}
    \includegraphics[scale=0.5]{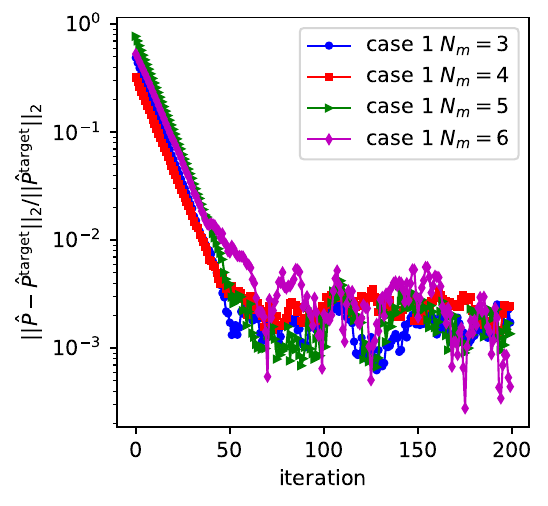} &
     \includegraphics[scale=0.5]{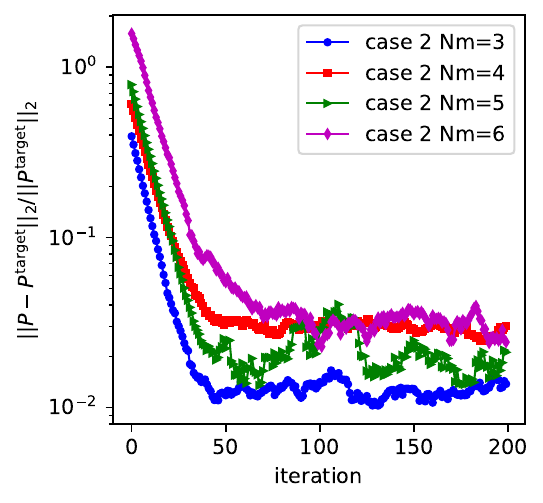} &
         \includegraphics[scale=0.5]{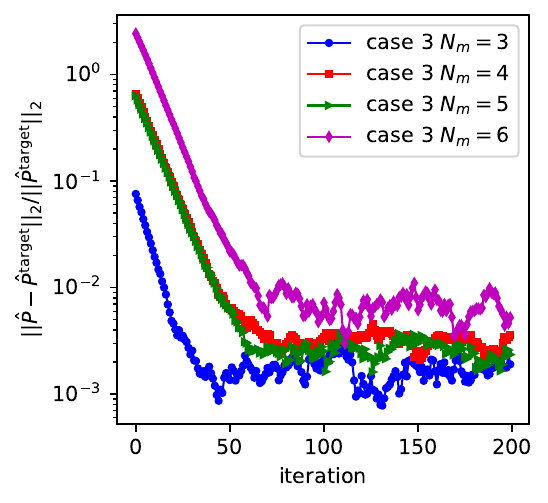} 
         \\
    \includegraphics[scale=0.5]{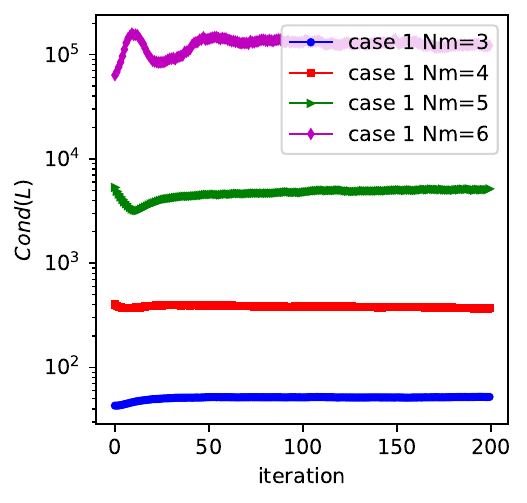}
   &
    \includegraphics[scale=0.5]{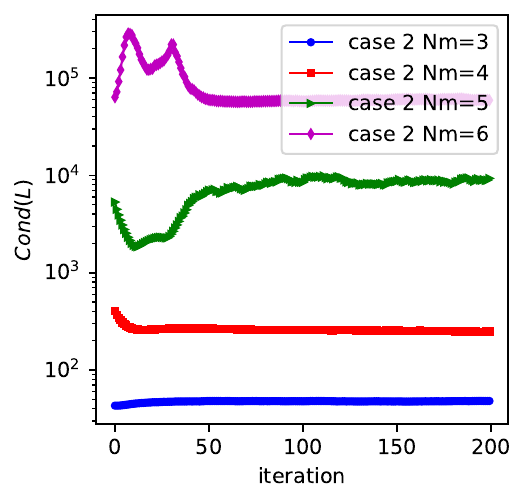}
    &
    \includegraphics[scale=0.5]{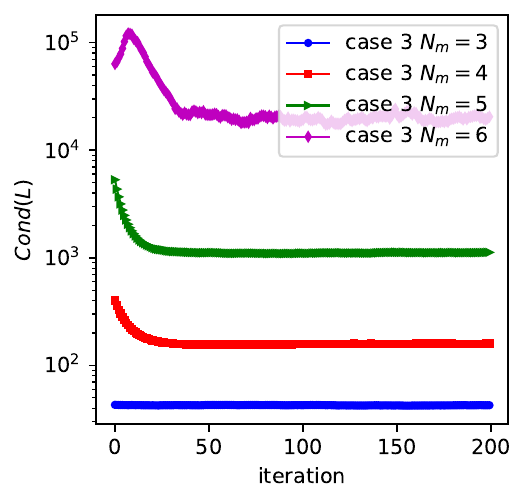}
    \\
 Case 1 & Case 2 & Case 3
       \end{tabular}
    \caption{The evolution of relative error in $N_m$ moments corresponding the polynomials $H^v$ and the condition number in WE realization of cases 1-3 of the bi-modal distribution (see eq.~\eqref{eq:bi-modal_dist} and Table~\ref{tab:param_bi-modal-dist}).}
    \label{fig:approx_bi_modal_err_cond}
\end{figure}

Since bi-modal distributions are often observed in high Mach-number flows, 
 here we investigate the accuracy of the proposed WE closure in realizing the target bi-modal distribution 
 \begin{eqnarray}
 f^\mathrm{target} &=&\alpha \mathcal{M}_v(1,\mu_1,\sigma_1)+(1-\alpha)\mathcal{M}_v(1,\mu_2,\sigma_2)
\label{eq:bi-modal_dist}
\end{eqnarray}
with $\alpha=0.5$, $\mu_2 = - \mu_1$ and $\sigma_2=\sqrt{2 - (\sigma_1 ^ 2 + 2  \mu_1 ^ 2)}$ which leads to zero mean and variance of unity. In particular, we consider the three cases summarized in Table~\ref{tab:param_bi-modal-dist}.
\begin{table}
    \centering
    \begin{tabular}{c|cc}
        Case & $\mu_1$ & $\sigma_1$ \\ \hline
        1 &  $0.8$ & $0.3$ 
        \\
        2 & $0.9$ & $0.1$
        \\
        3 & $0.9$ & $0.4$
    \end{tabular}
    \caption{Parameters of bi-modal distributions}
    \label{tab:param_bi-modal-dist}
\end{table}
\noindent Given the moments of the target bi-modal distribution \eqref{eq:bi-modal_dist}, we use algorithm \ref{alg:WE} to find samples of the WE closure as well as the corresponding Lagrange multipliers. We deploy $10^4$ particles and take $\tau/dt=10$ and iterate the WE algorithm for $200$ steps after which no discernible evolution takes place.
\\ \ \\
In Fig.~\ref{fig:bi-mod-distr} we present the WE closure solution obtained using the proposed stochastic solution algorithm by matching moments of the polynomials $H^v=\{v,v^2,...,v^{N_m}
\}$ where $N_m\in \{ 3,4,5,6\}$. Similar to previous sections, for comparison we also report the MED distribution computed using a Gaussian prior within a maximum cross-entropy formulation \cite{debrabant2017micro}  with $N_m=4,6$. The figure shows reasonable qualitative agreement between the WE and MED solutions.
\\ \ \\
Fig.~\ref{fig:approx_bi_modal_err_cond} shows the evolution of relative error $||\hat P-\hat P^{\mathrm{target}}||_2/||\hat P^{\mathrm{target}}||_2$ where $\dim( \hat P ) = N_m$  and the condition number of the matrix in eq.~\eqref{eq:matrix_to_invert}. Even though the condition number increases with the order of moments $N_m$, the particle solution algorithm presented in Algorithm~\ref{alg:WE} still provides a reasonable relative error in the moment matching procedure. The WE solution exists for all values of $N_m$ with good accuracy compared to the ideal MED solution, where the latter exists.

\section{Application to direct Monte Carlo computations}
\label{sec:results}
As outlined in the introduction, the proposed closure has potential applications to particle simulations of transport. In the sections that follow, we use DSMC computations to assess the ability of the proposed method to fulfill that role.
\\ \ \\
In the simulations that follow we use the hard-sphere model for argon with hard-sphere diameter $d=3.405\times 10^{-10}\ \mathrm{m}$ and molecular mass  $m=6.6335214 \times 10^{-26}\ \mathrm{kg}$, leading to a mean free path $\tilde{\lambda}(n_0)=(\sqrt{2}\pi d^2 n_0 )^{-1}$, diameter. In all simulations, we resolve the mean free time as well as the traversal time using $\Delta t = 0.5 \textrm{min} \left( \tilde{\lambda}, \Delta x_2 \right)/\textrm{max}(\sqrt{k_bT_0/m},U^{\mathrm{wall}})$.

\subsection{Re-sampling DSMC computations}
\label{sec:resampling_DSMC}
  In this section, we use our methodology to initialize DSMC computations using a number of moments obtained from the DSMC simulations themselves. In other words, at a given time, specified below for each problem studied, we sample the DSMC simulation to obtain an estimate of its moments, up to and including the heat flux, namely, 
  \begin{flalign}
      H= \left[ 1,\ v_1,\ v_2,\ v_3,\ \xi_1^2,\ \xi_1 \xi_2,\ \xi_1 \xi_3,\ \xi_2^2,\ \xi_2 \xi_3,\ \xi_3^2,\ \xi_1 \left(\sum_{i=1}^3 \xi_i^2\right),\ \xi_2 \left(\sum_{i=1}^3 \xi_i^2\right),\ \xi_3  \left(\sum_{i=1}^3 \xi_i^2 \right) \right]
  \end{flalign}
where $\xi = v - U$ is the fluctuating velocity.  We then use these moments to initialize a DSMC computation using our proposed methodology. The discrepancy between a reference, unperturbed, DSMC computation and the initialized DSMC computation for a number of canonical problems is then used as a measure of the effectiveness of our proposed approach.   
\subsubsection{Standing wave}
\label{sec:standing_shock}
Let us consider a simulation of a standing wave \cite{lockerby2008modelling}, which avoids the issue of initializing in the vicinity of solid walls, where non-equilibrium effects are expected to be stronger. In other words, this case investigates the proposed method's ability to initialize DSMC computations in bulk.
\\ \ \\
We simulate the evolution of gas particles between $x_2\in[0,L]$ with periodic boundary conditions in the presence of a harmonic external body force 
\begin{eqnarray}
F_1 &=& A \cos( \alpha t ) \cos(\beta \frac{2 \pi x_2}{L})
\end{eqnarray}
 where $A=10^6\ \mathrm{kg.m.s^{-2}}$,  mode number $\beta=1$, and frequency $\alpha=0.1 / \tilde{t}$; the latter is computed using the mean free time $\tilde{t}=\tilde{\lambda}(n_0)/\sqrt{\theta(T_0)}$. We study the solution for a wide range of   $\textrm{Kn}\in\{1,0.1,0.01\}$ by changing the distance $L$. We resolve the mean free path by considering a cell size of $\tilde{\lambda}/100, \tilde{\lambda}/10,$ and $\tilde{\lambda}/2$.   Here we deploy on average $1000$ particles per cell and estimate moments using $200$ ensembles. At $t=0$, particles are initialized from the Maxwell-Boltzmann equilibrium distribution function with $T(x,t=0)=T_0=273\ \mathrm{K}$, bulk velocity $U(x,t=0)=0$, and number density $n(x,t=0)=n_0=10^{20}\ \mathrm{m}^{-3}$.
 \\ \ \\
 Given this is a transient problem, and in order to assess the possibility of error accumulation from our proposed method, the initialization process is repeated at regular intervals of 100 steps. In other words, every 100 steps, particle velocities are re-initialized using algorithm \ref{alg:WE} with moments taken from the same calculation at that time.  
 \begin{figure}
    \centering
    \begin{tabular}{ccc}   \includegraphics[scale=0.55]{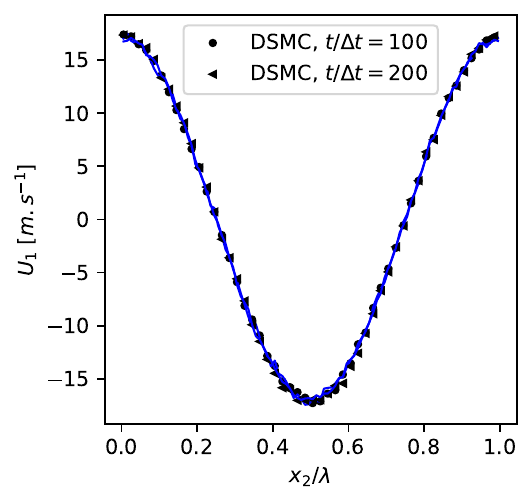}
    &
    \includegraphics[scale=0.55]{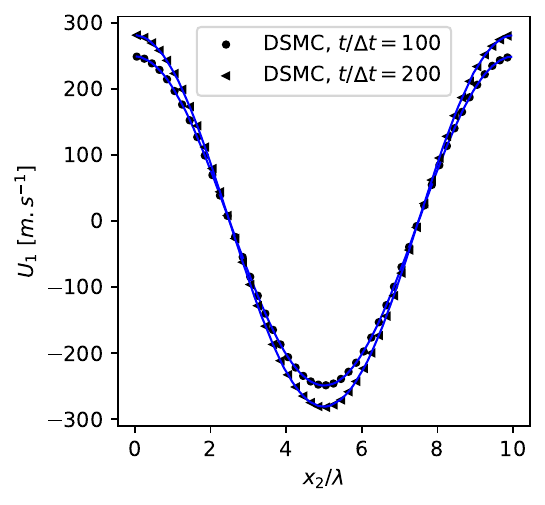}
    &
    \includegraphics[scale=0.55]{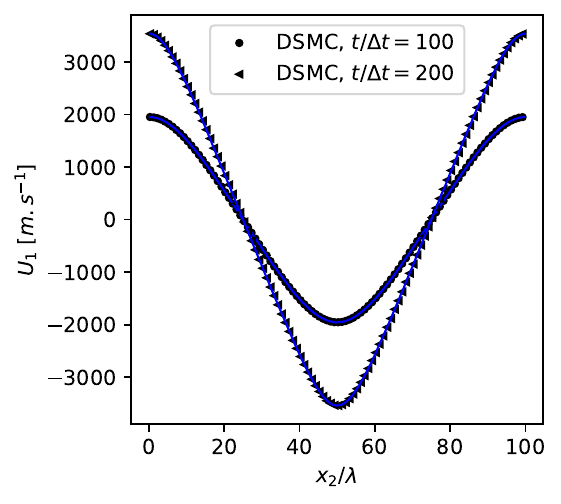}
    \\
      \includegraphics[scale=0.55]{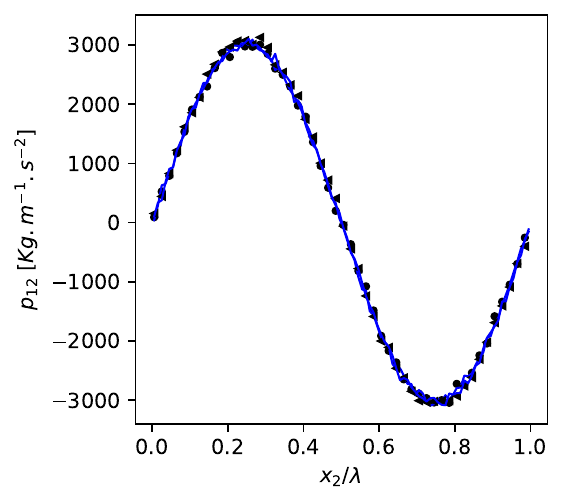}
      &
       \includegraphics[scale=0.55]{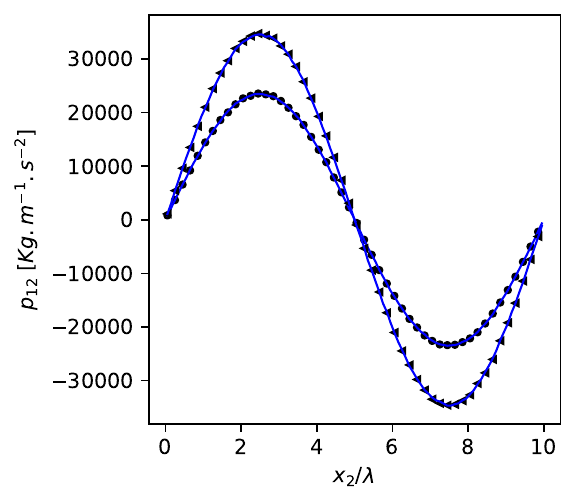}
       &
        \includegraphics[scale=0.55]{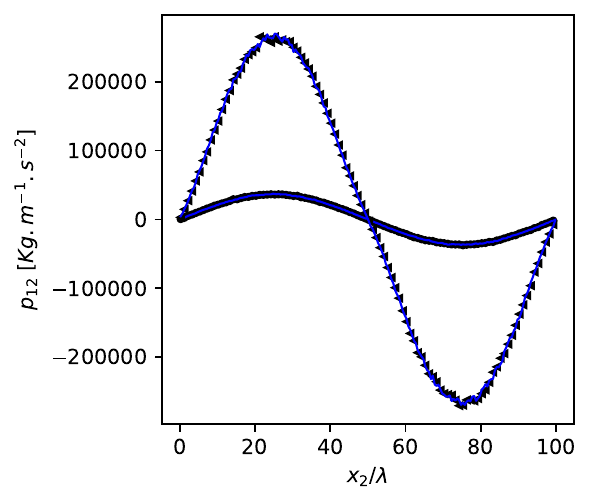}
\\
      \includegraphics[scale=0.55]{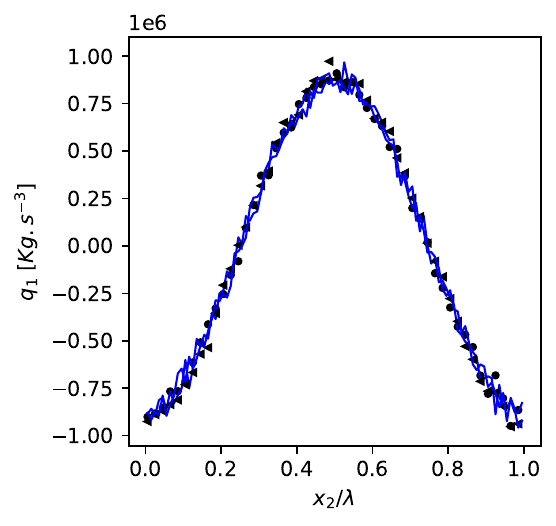}
      &
      \includegraphics[scale=0.55]{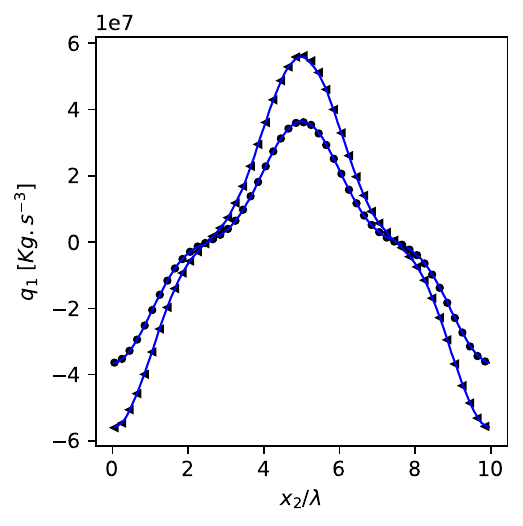}
      &
      \includegraphics[scale=0.55]{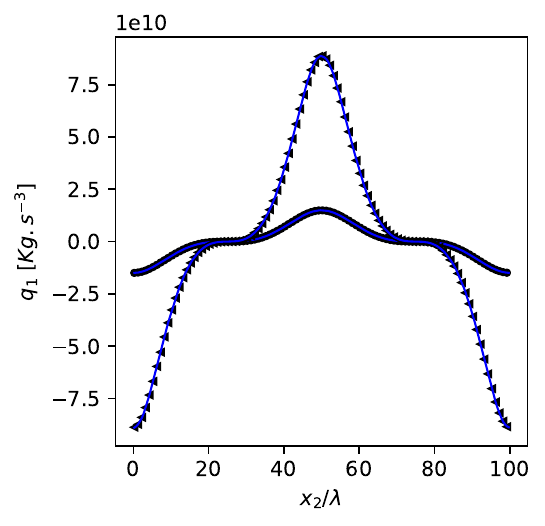}
      \\
      $\mathrm{Kn}=1$ & $\mathrm{Kn}=0.1$ & $\mathrm{Kn}=0.01$
    \end{tabular}
    \caption{Transient solution of bulk velocity, shear stress, and heat flux to the standing wave problem for $\mathrm{Kn}=1,0.1,0.01$ at $t/\Delta t \in \{100, 200\}$
    using standard DSMC (black) and resampled DSMC every $100$ steps using WE closure matching up to heat flux (blue). }
\label{fig:standing_wave_results}
\end{figure}
Figure \ref{fig:standing_wave_results} shows a comparison between this transient solution and the reference  (unperturbed) DSMC computation. The figure shows that the re-initialized simulation follows the benchmark DSMC solution with high accuracy, as expected from the results of Sec. \ref{sec:MomentSystemLimit}, where it was shown that providing information up to the heat flux results in initialization equivalent to a Navier-Stokes-Fourier (NSF) level of description. This makes the very good agreement at $\mathrm{Kn}=1$ particularly encouraging.  In the next section, we further test the accuracy of the WE approach in the presence of solid boundaries which are known to introduce additional kinetic effects  \cite{hadjiconstantinou2006limits,sone2007molecular}.
\subsubsection{Couette Flow}
\label{sec:couette_flow}
\begin{figure}
    \centering
    \begin{tabular}{ccc}
    \includegraphics[scale=0.55]{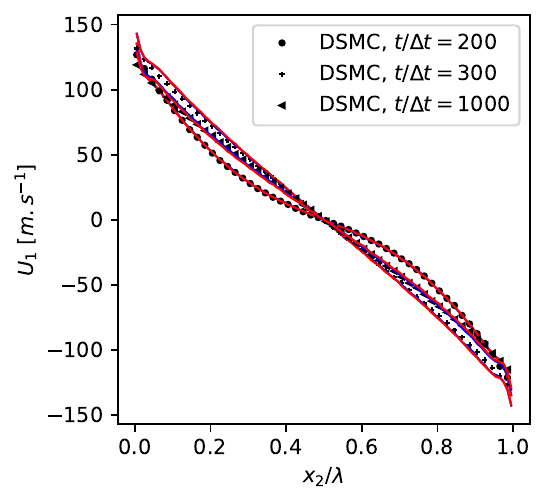}
    &
    \includegraphics[scale=0.55]{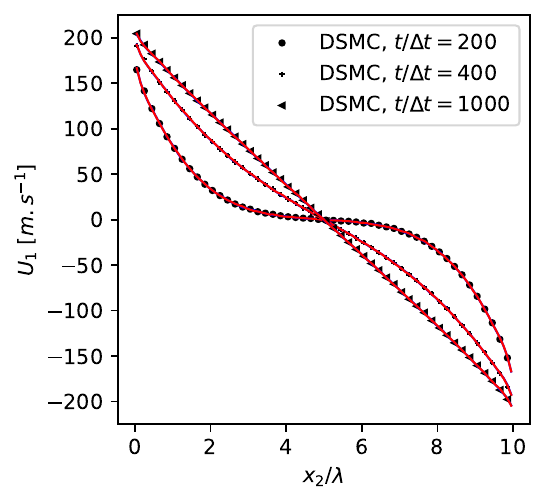}
    &
    \includegraphics[scale=0.55]{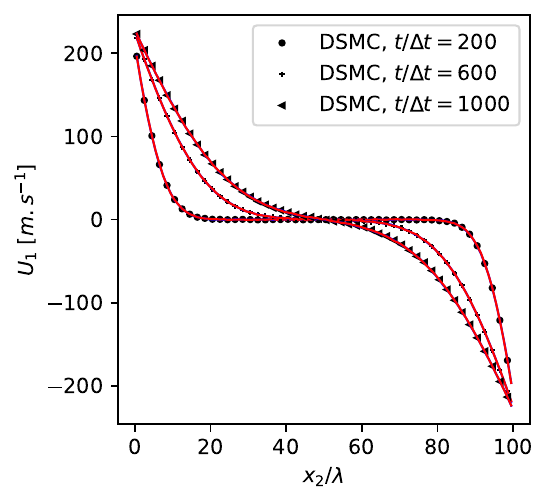}
    \\
    \includegraphics[scale=0.55]{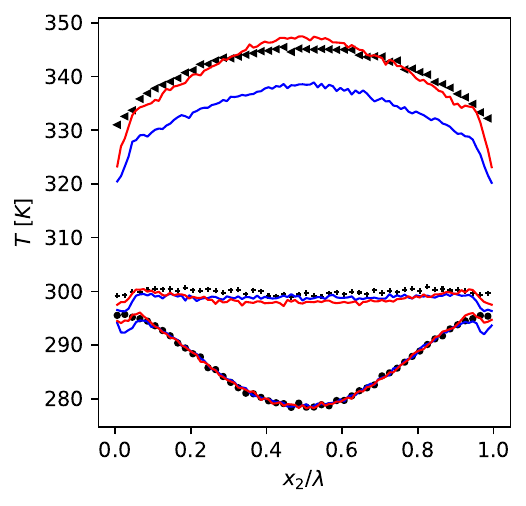}
    &
    \includegraphics[scale=0.55]{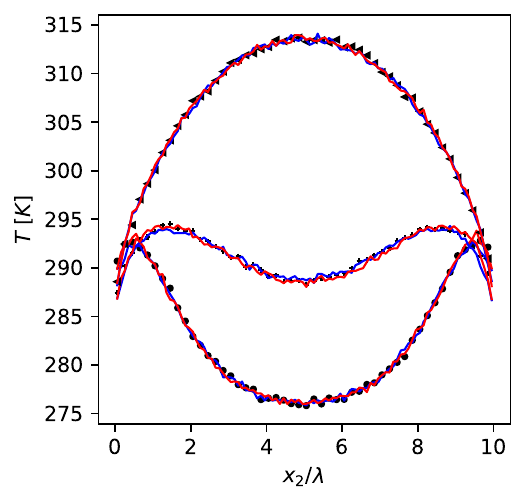}
    &
    \includegraphics[scale=0.55]{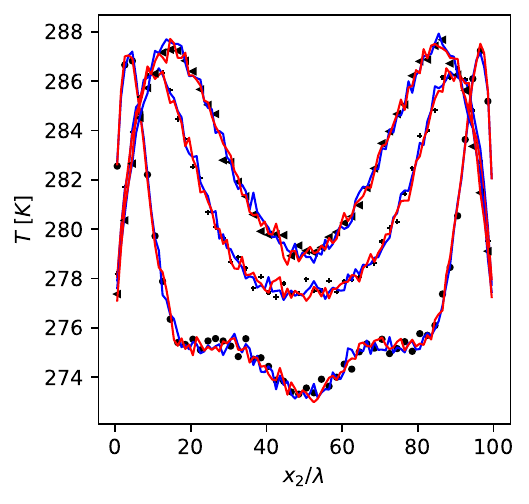}
    \\
    \includegraphics[scale=0.55]{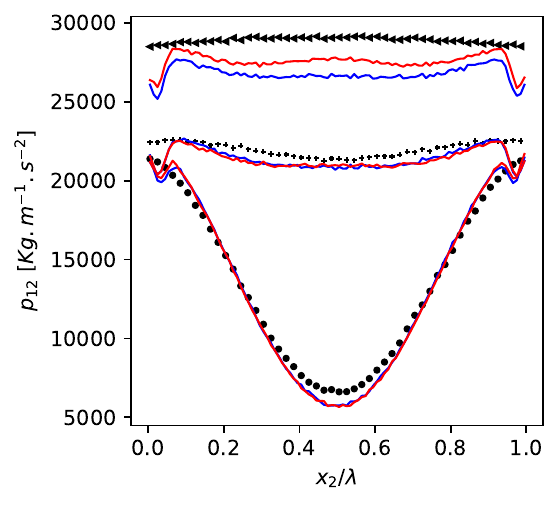} &
    \includegraphics[scale=0.55]{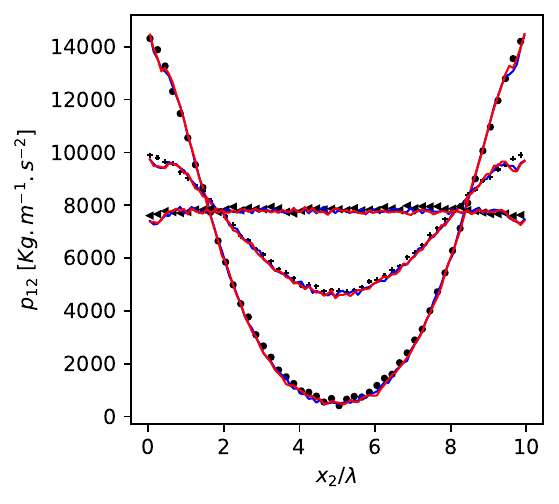}
    &
    \includegraphics[scale=0.6]{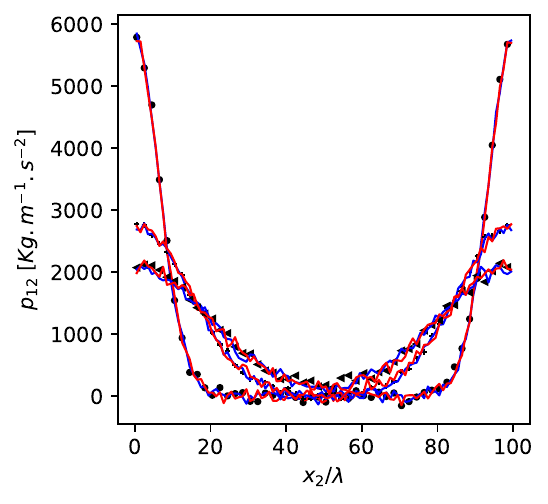}
    \\
    \includegraphics[scale=0.55]{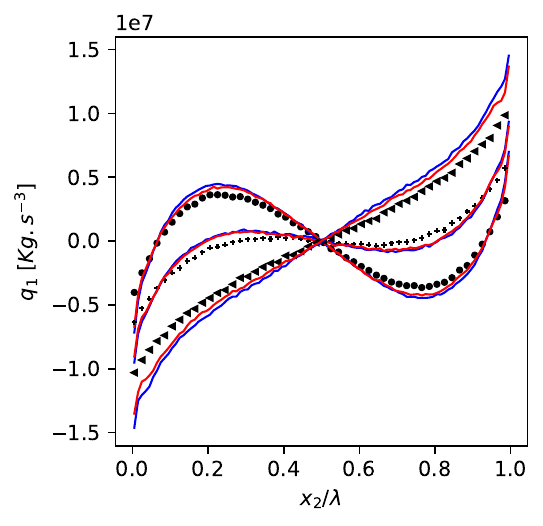}
    &
    \includegraphics[scale=0.55]{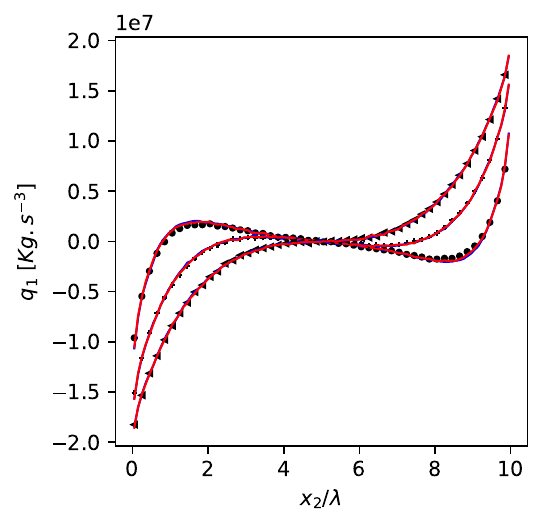}
    &
    \includegraphics[scale=0.55]{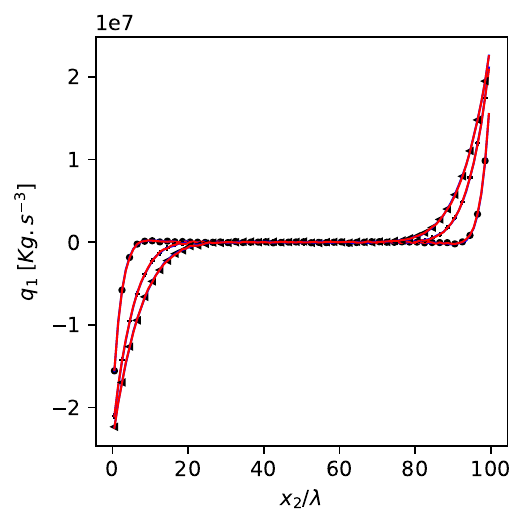}
      \\
      $\mathrm{Kn}=1$ & $\mathrm{Kn}=0.1$ & $\mathrm{Kn}=0.01$
    \end{tabular}
    \caption{Transient solution for the bulk velocity, temperature, shear stress, and heat flux in a $\mathrm{Ma}=1$ Couette flow for three values of the Knudsen number. Comparison between standard DSMC (black) and DSMC with resampling every $100$ steps using the WE closure matching up to heat flux (blue) and up to 4th order moment (red). Solutions are shown at $t/\Delta t \in \{200, 300, 1000\}$, $\{200,400,1000\}$, $\{200,600,1000\}$ for $\mathrm{Kn}=1,0.1,0.01$, respectively. }
\label{fig:resample_Couette}
\end{figure}

\begin{figure}
    \centering
    \begin{tabular}{cc}
    \includegraphics[scale=0.6]{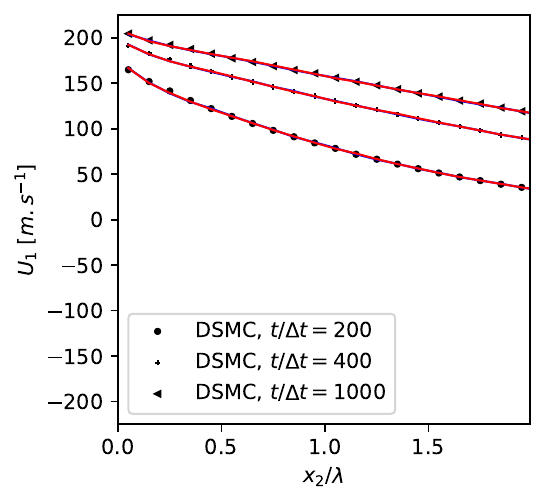}
    &
    \includegraphics[scale=0.6]{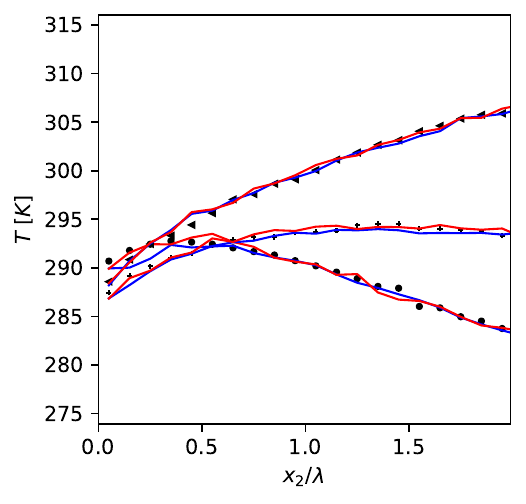}
    \\
    \includegraphics[scale=0.6]{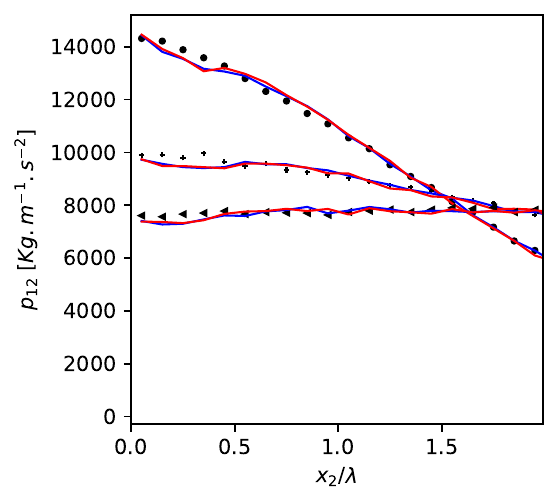} &
    \includegraphics[scale=0.6]{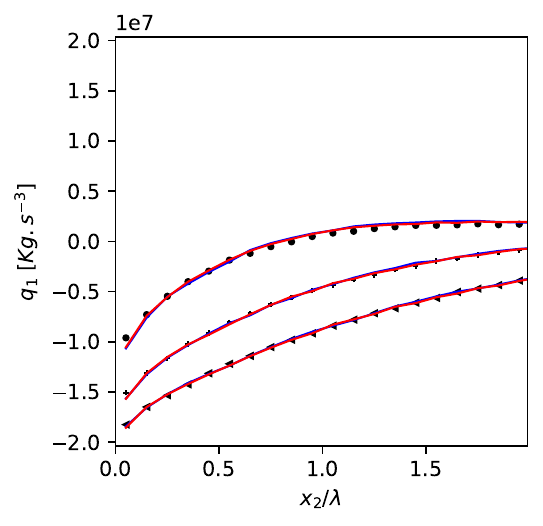}
    \end{tabular}
    \caption{Detail of hydrodynamic profiles in the wall vicinity for the $\mathrm{Ma}=1$, $\mathrm{Kn}=0.1$ Couette flow discussed in Fig. \ref{fig:resample_Couette}. Standard DSMC is shown in black, while resampled DSMC  using moments up to heat flux and up to 4th order moment are shown in blue and red, respectively.}
    \label{fig:resample_Couette_kn0.1_zoomed}
\end{figure}

In this section, we present results from the transient simulations of argon gas in a one-dimensional Couette flow problem, We investigate the performance of the WE method at a wide range of   $\textrm{Kn}\in\{1,0.1,0.01\}$ by changing the distance $L$ between walls.  The boundaries located at $x_2=0$ and $x_2=L$ are thermal walls with temperature $T^\mathrm{wall}=273\ \mathrm{K}$ and  velocities $U^\mathrm{wall}=\pm \mathrm{Ma} \sqrt{k_bT_0/m}$, where $\mathrm{Ma}$ denotes the Mach number. 
 The mean free path is resolved by considering a cell size of $\tilde{\lambda}/100, \tilde{\lambda}/10, \tilde{\lambda}/5$ and $\tilde{\lambda}/2$. We again deploy on average $1000$ particles per cell and estimate the moments using $1000$ ensembles.
\\ \ \\
 Fig.~\ref{fig:resample_Couette} shows a comparison for $\textrm{Ma}=1$, which follows the same comparison protocol as the standing-wave problem of the previous section. In other words, a standard transient DSMC computation is compared with a DSMC computation resampled every 100 timesteps with moments taken from the perturbed solution at the resampling time. 
The figure shows, as expected perhaps, that the WE closure works very well for $\mathrm{Kn} \ll 1$, while error is clearly visible for $\mathrm{Kn}=1$. Moreover, small error is visible in the wall vicinity, within one mean free path distance from the walls for $\mathrm{Kn} \ll 1$ (see Fig. \ref{fig:resample_Couette_kn0.1_zoomed} for a detailed comparison). This error is attributed to the Knudsen layer contributions which are not described by the Chapman-Enskog distribution.
\\ \ \\
As one would expect, the discrepancy observed at $\mathrm{Kn}=1$ can be rationalized by the importance of higher order moments as $\mathrm{Kn}$ increases beyond the NSF limit and motivates the inclusion of such moments in the WE closure. To this end, in
Figure \ref{fig:resample_Couette}-\ref{fig:resample_Couette_kn0.1_zoomed} we also show the results of a comparison in which the WE procedure makes use of  moments up to 4th order, by including the polynomials
\begin{flalign}
    \left[ \xi_1^2 \left(\sum_{i=1}^3\xi_i^2\right),\ \xi_2^2  \left(\sum_{i=1}^3\xi_i^2\right),\ \xi_3^2  \left(\sum_{i=1}^3\xi_i^2\right)~ \right] . \nonumber
\end{flalign}
The clear improvement in the results compared to the closure that only matches up to heat flux illustrates the dependence of the solution on higher-order moments. We leave a more detailed investigation to future work.
\subsection{Resuming DSMC solution at steady state}
\begin{figure}
    \centering
    \begin{tabular}{ccc}
    \includegraphics[scale=0.5]{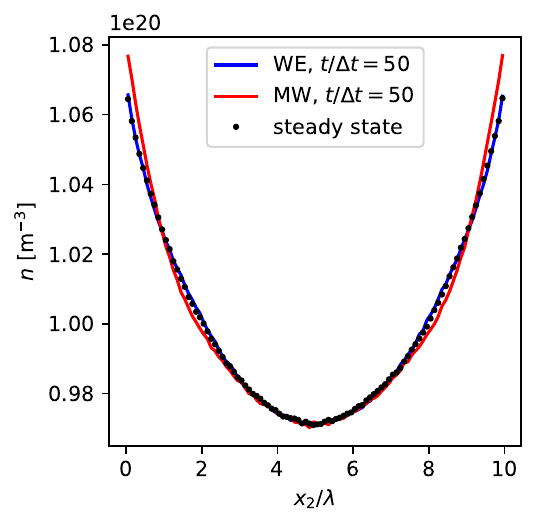}
    &
    \includegraphics[scale=0.5]{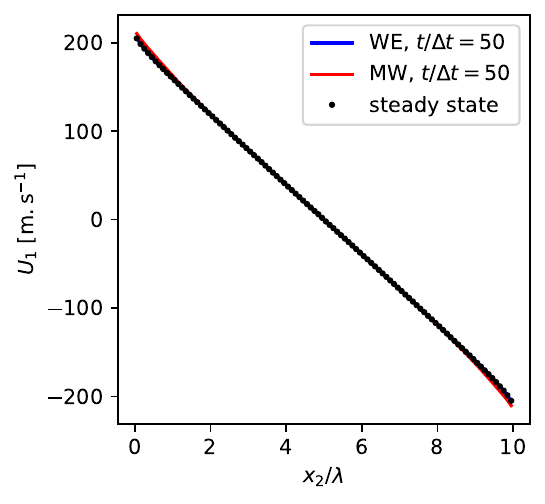}
    &
    \includegraphics[scale=0.5]{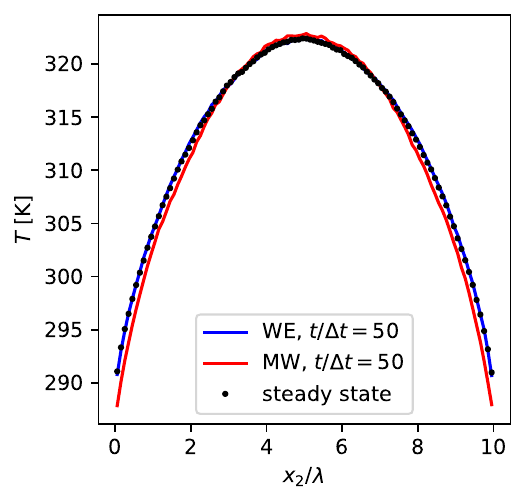}
    \\
    \includegraphics[scale=0.5]{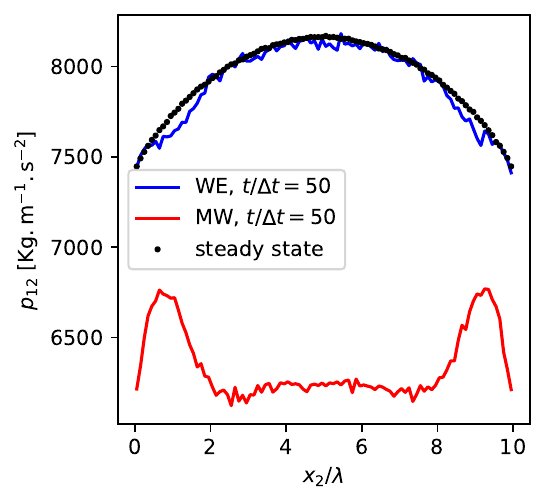}
    &
    \includegraphics[scale=0.5]{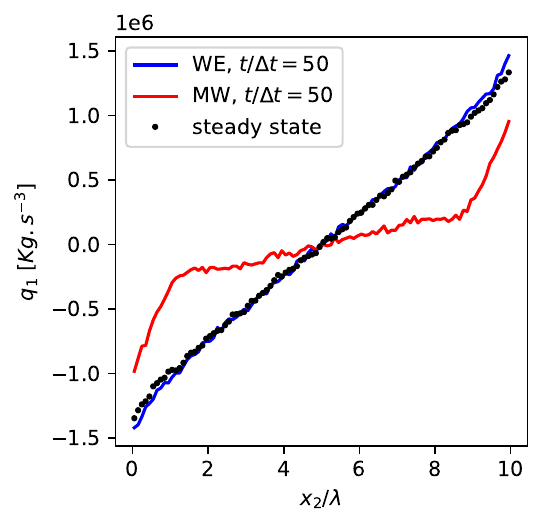}
    &
    \includegraphics[scale=0.5]{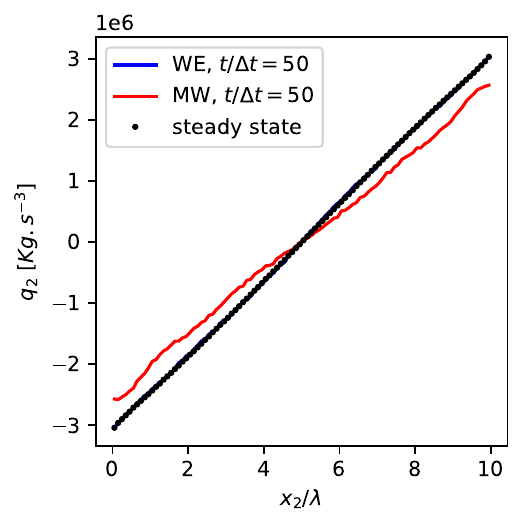}
    \end{tabular}
    \caption{Resuming steady state solution of DSMC (black) for the Couette flow at $\mathrm{Kn}=0.1$ using WE (blue) local Maxwellian (red) with  $10^4$ ensembles.}
\label{fig:resample_steady-Couette_kn0.1}
\end{figure}

\begin{figure}
    \centering
    \begin{tabular}{ccc}
    \includegraphics[scale=0.5]{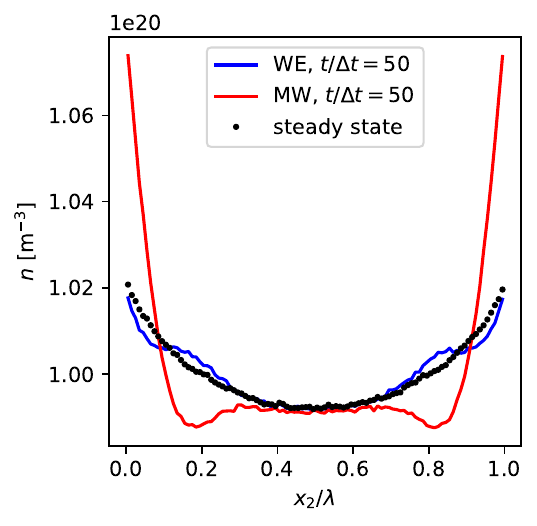}
    &
    \includegraphics[scale=0.5]{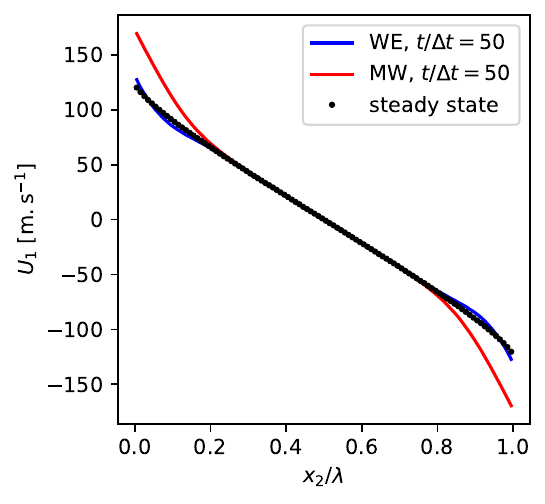}
    &
    \includegraphics[scale=0.5]{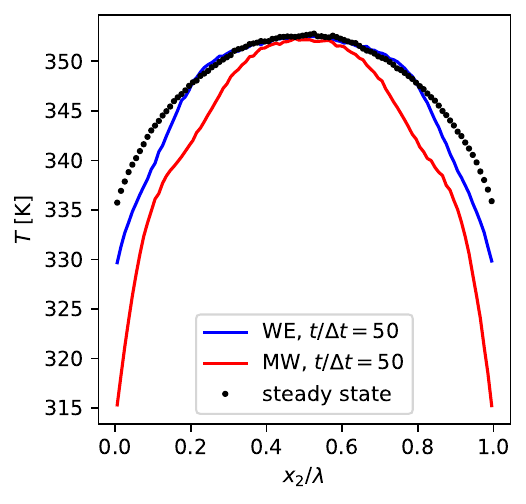}
    \\
    \includegraphics[scale=0.5]{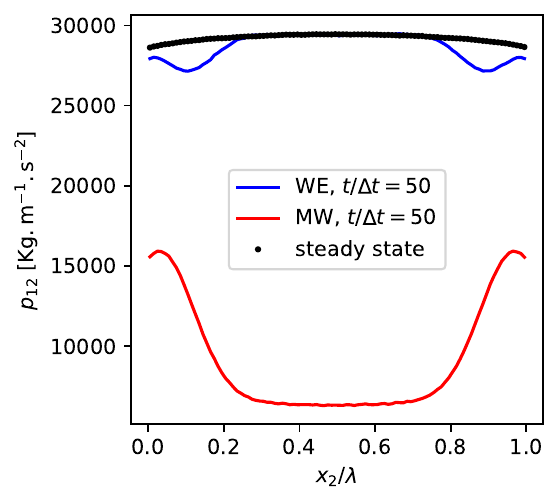}
    &
    \includegraphics[scale=0.5]{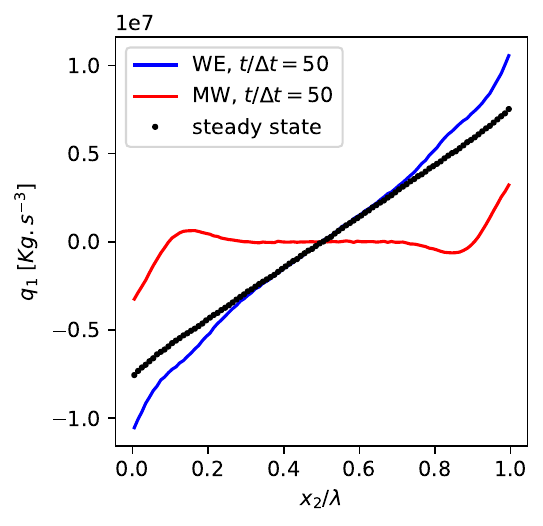}
    &
    \includegraphics[scale=0.5]{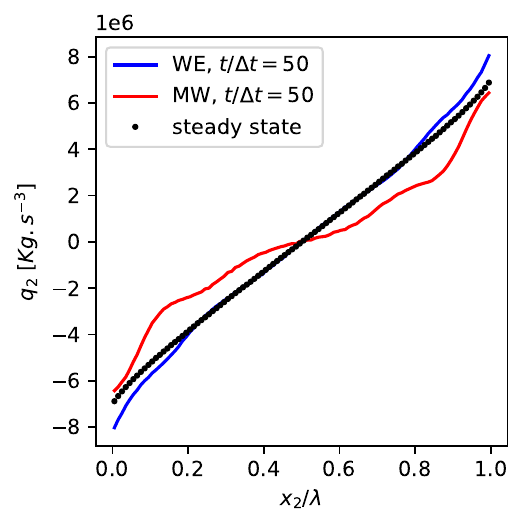}
    \end{tabular}
    \caption{Resuming steady state solution of DSMC (black) for the Couette flow at $\mathrm{Kn}=1$ using WE (blue) local Maxwellian (red) with  $10^4$ ensembles.}
\label{fig:resample_steady-Couette_kn1}
\end{figure}

Here, we further examine the proposed method's ability to create samples of an underlying microscopic velocity distribution function given macroscopic information. With this test case, we assess the possibility of using the proposed method within a steady-state solution framework as described, for example. in the equation-free methodology  \cite{theodoropoulos2000coarse}. 
\\ \ \\
As one would expect, any error from re-initializing DSMC simulations as part of a root-finding iteration process will manifest itself as steady-state error \cite{al2007acceleration}. To simplify the computation, we investigate this error using the steady solution as a starting point; we expect any initialization error to cause the initialized simulation to move away from the correct solution.
\\ \ \\
Figures 
\ref{fig:resample_steady-Couette_kn0.1}-\ref{fig:resample_steady-Couette_kn1}, compare steady-state DSMC results obtained after initializing from the steady DSMC solution using the local equilibrium distribution function and the WE method for two different Knudsen numbers; the WE initialization uses moment information up to the heat flux. The steady solution serves as a reference from which deviations are measured. As expected, the DSMC computation initialized using the WE process is significantly  closer to the steady-state solution than the one sampled using local equilibrium.  This result provides further evidence that the  WE method can be used to enable particle method acceleration schemes such as the equation-free framework \cite{theodoropoulos2000coarse}.

\section{Conclusion}
\label{sec:conclusion}
\noindent In this work we present a new closure to the problem of generating samples from a distribution identified only by a small number of its moments. The proposed closure, referred to as Wasserstein-penalized Entropy, combines minimization of entropy with the Wasserstein distance from an input, auxiliary, distribution, usually taken as the local equilibrium. This leads to a well-defined distribution for the entire space of realizable moments.  We also developed an efficient Monte Carlo solution algorithm for generating samples of the target distribution. We expect the Monte Carlo formulation to lend itself naturally to high-dimensional settings where more deterministic approaches typically suffer.
\\ \ \\
We demonstrate that in the case of realizable target moments this solution algorithm converges monotonically and provides samples matching target moments within statistical noise. We further show that in the case of non-realizable (non-physical) moments, the method can be stopped in the realizable neighbourhood of target moments given the monotone convergence of the proposed time-stepping method. We show analytically that the proposed closure recovers the Euler and Navier-Stokes-Fourier equations in the hydrodynamic limit while maintaining a well-defined distribution function. Furthermore, in several numerical studies on prototypical internal flow problems, we observe that the proposed closure can reasonably approximate the solution of the Boltzmann equation for $\mathrm{Kn} \ll 1$, by relying on moment information only up to the heat fluxes. Extension into further rarefied regimes requires the use of higher-order moments, which can be achieved in a straightforward way. 
\\ \ \\
We also emphasize that the proposed approach offers direct samples of the target probability density by leveraging the introduced SDE representation. This enables the samples to explore the entire phase space, especially important when dealing with high Mach flows. The resulting computational advantage is particularly notable as conventional closure methods, when employed in stochastic particle systems, require separate treatments of the sampling problem. The latter often entails restrictions on the sample space and sub-optimal scaling with the number of dimensions.  \\ \ \\
We anticipate that the proposed methodology will facilitate computational techniques which integrate atomistic and continuum-based methods for solving multiscale problems. In addition, the proposed method is expected to have applications more broadly to particle simulation methods, including topics such as variance reduction and particle-number control \cite{gorji2019particle}. 

\section*{Acknowledgments}
\noindent 
 MS acknowledges the funding provided by the German research foundation (DFG) under grant number SA 4199/1-1. 
\appendix
\section{Regularity of WE}
\label{proof:L1}
\noindent In the following we provide justification for $\pi\in K_\pi$. Conceptually since the Wasserstein term goes to $-\infty$ with a power larger than the polynomials $H$ considered in the maximum entropy part, the exponential goes to zero as $(v^2+w^2)\to \infty$. The details are provided in the justification of the following proposition. 
\begin{prop}
\label{prop-L1}
Suppose the polynomial basis $H(v,w)$ grows by the exponent $k$ at infinity $(k\in\mathbb{N})$. Let $p=k+1$. For $0<\alpha<1$, $0<C_0<\infty$, and finite $\lambda$ we have
\begin{eqnarray}
\label{eq:L1-const}
\int_{\mathbb{R}^3\times\mathbb{R}^3}\exp\left(\lambda_i H_i-{\alpha} C_0|v-w|^p \right) \ \textrm{\textup{dvdw}}&<& +\infty \ .
\end{eqnarray}
\end{prop}
\begin{proof}
For simplicity and without loss of generality let us consider the scenario where $H$ grows by the exponent $2k+1$ at infinity with a positive prefactor. Thus it would be sufficient to show 
\begin{eqnarray}
I&=&\int_{\mathbb{R}^3\times\mathbb{R}^3}\exp\left(C_1|v|^{2k+1}+C_2|w|^{2k+1}-C_3|v-w|^{2k+2} \right) \ \text{dvdw}
\end{eqnarray}
is finite, for $0<C_{1,2,3}<\infty$. However notice
\begin{eqnarray}
I&\leq&J=\int_{\mathbb{R}^3\times\mathbb{R}^3}\exp\left(C_1|v|^{2k+1}+C_2|w|^{2k+1}-C_3\left|\ |v|^{2k+2}-|w|^{2k+2}\ \right| \right) \ \text{dvdw},
\label{eq:J}
\end{eqnarray}
due to the reverse triangle inequality. The upper bound $J$ can be split into two integrals
\begin{eqnarray}
J&=&J_1+J_2,\\
\label{eq:J1}
J_1&=&\int_{\mathbb{R}^3\times\mathbb{R}^3}{\bm{\mathds{1}}}_{|v|>|w|}\exp\left(C_1|v|^{2k+1}+C_2|w|^{2k+1}+C_3|w|^{2k+2}- C_3|v|^{2k+2} \right) \ \text{dvdw}
\end{eqnarray}
and
\begin{eqnarray}
\label{eq:J2}
J_2&=&\int_{\mathbb{R}^3\times\mathbb{R}^3} {\bm{\mathds{1}}}_{|w|>|v|}\exp\left(C_1|v|^{2k+1}+C_2|w|^{2k+1}+ C_3|v|^{2k+2}-C_3|w|^{2k+2} \right) \ \text{dvdw},
\end{eqnarray}
where $\bm{ \mathds{1}}$ is the indicator function. Next, let us translate the integrals of $J_1$ and $J_2$ into the  six-dimensional sphere with the radius $r^2=|v|^2+|w|^2$ and the angles $\psi=[\psi_1,...,\psi_{5}]$, where $\psi_{1,...,{4}}\in [0,\pi]$ and $\psi_{5}\in [0,2\pi)$. Let the Jacobian of the transformation be $r^{5}l_0(\psi)$ (note that $l_0(\psi)=\sin^{4}(\psi_{1})\sin^3(\psi_2)\sin^2(\psi_3)\sin(\psi_{4})$, see e.g. \cite{blumenson1960derivation}). Furthermore, suppose $|v|=r|l_v(\psi)|$ and $|w|=r|l_w(\psi)|$. Hence the integrals can be put in the following forms
\begin{eqnarray}
J_1&=&\int_{\psi}\overbrace{\int_0^\infty {\bm{\mathds{1}}}_{|l_v(\psi)|>|l_w(\psi)|}\exp\left(A_\psi r^{2k+1}-B_\psi r^{2k+2} \right) r^{5}l_0(\psi)\ \text{dr}}^{F_1(\psi)} \text{d$\psi$}, 
\end{eqnarray}
and similarly 
\begin{eqnarray}
J_2&=&\int_{\psi}\overbrace{\int_0^\infty  {\bm{\mathds{1}}}_{|l_w(\psi)|>|l_v(\psi)|}\exp\left(A_\psi r^{2k+1}-B_\psi r^{2k+2}\right)r^{5}l_0(\psi)\ \text{dr}}^{F_2(\psi)} \text{d$\psi$}, 
\end{eqnarray}
where the positive pre-factors 
\begin{eqnarray}
A_\psi&=&C_1 |l_v(\psi)|^{2k+1}+C_2 |l_w(\psi)|^{2k+1} \\
\textrm{and} \ \ \ \ B_\psi&=&C_3\left||l_v(\psi)|^{2k+2}-|l_w(\psi)|^{2k+2}\right|
\end{eqnarray}
only depend on $\psi$. 
The finiteness of $J_1$ and $J_2$ (and thus $J$ which is sum of the two) can be justified if the inner most integral, i.e. the integral with respect to $r$, is finite. This is due to the fact that the integration domain of the other integral, namely the integral with respect to $\psi$, is bounded. In other words if $F_1(\psi)$ and $F_2(\psi)$ are finite for all values of $\psi_{1,...,{4}}\in [0,\pi]$ and $\psi_{5}\in [0,2\pi)$, $J_1$ and $J_2$ will be finite. However, the inner most integrals (and hence $F_1(\psi)$ and $F_2(\psi)$) are finite since the exponential function decays with exponent $r^{2k+2}$, as $r\to\infty$, and thus the condition \eqref{eq:L1-const} is satisfied. 
\end{proof}
\noindent Similarly, we can show that $\langle H \rangle_\pi$ is finite, as $\exp(-B_\psi r^{2k+2})$ goes to zero faster than polynomials. Consequently, we have $\pi\in K_\pi$.

\noindent 

\section{Navier-Stokes-Fourier Limit}
\label{proof:NSF}
\noindent Following \cite{levermore1996moment},  consider the polynomials $H_{i\in \{1,...,14\}}^v$ which belong to 
\begin{eqnarray}
\mathbb{M}^{v}&=&\textrm{span}\left\{1,v_1,v_2,v_3,,v_1v_1, v_1v_2,v_1v_3,v_2v_2, v_2v_3, v_3v_3,|v|^2v_1,|v|^2v_2,|v|^2v_3, |v|^4\right\}\ .
\end{eqnarray}
We note that we included $|v|^4$ in $\mathbb{M}^{v}$ to have a similar setting as the one of Levermore \cite{levermore1996moment}. However, similar results can be obtained using polynomials up to third order for WE.
\\ \ \\
\noindent Let us define the linearized collision operator 
\begin{eqnarray}
\mathscr{L}_{f^{(0)}}\bigg[\phi (v)\bigg]&=&-\frac{1}{f^{(0)}}\partial_s\mathcal{C}\left[f^{(0)}(1+s\phi(v))\right]\bigg|_{s=0} 
\end{eqnarray}
and assume that the collision operator $\mathcal{C}[.]$ admits the positivity property such that the matrix $\mathcal{D}$ defined by
\begin{eqnarray}
\label{eq:D}
\mathcal{D}_{ij}&=&\left \langle H_i^v \mathscr{L}_{f^{(0)}}[H_j^v]\right\rangle_{f^{(0)}}
\end{eqnarray} 
is positive definite  (see inequality (2.23) and page 1052 of \cite{levermore1996moment} for details and justifications). 
Therefore the moment system takes the form of 
\begin{eqnarray}
\label{eq:moment-WE-NS}
\partial_t \langle H^v_i \rangle_{f_0}+\partial_{x_j}\langle v_jH^v_i \rangle_{f_0}&=& \bigg\langle H^v_i\mathscr{L}_{f^{(0)}}\left[\delta f^{(1)}\right] \bigg\rangle_{f_0}
\nonumber \\
&=& \mathcal{D}_{ij} \lambda_j^{v(1)}
\end{eqnarray}
as $\epsilon\to 0$, where $H^v_i\in 
\mathbb{M}^v$. Here, we used Eq.~\eqref{eq:df1_Hlam}. 
\begin{prop}
Given positive definite $\mathcal{D}$, see Eq.~\eqref{eq:D}, the moment system \eqref{eq:moment-WE-NS} has the form of the Navier-Stokes-Fourier system, with the closures
\begin{eqnarray}
\Sigma_{ij}&=&-\mu \left(\frac{\partial u_i}{\partial x_j}+\frac{\partial u_j}{\partial x_i}-\frac{2}{3}\frac{\partial u_s}{\partial x_s}\delta_{ij}\right) 
\end{eqnarray}
and
\begin{eqnarray}
q_i&=&-\kappa \frac{\partial \theta}{\partial x_i}
\end{eqnarray}
where $\mu,\kappa >0$.
\end{prop}
\begin{proof} 
Note that the Chapman-Enskog expansion of the Boltzmann collision operator gives us the following constraint  
\begin{eqnarray}
\label{eq:NSF1}
\bigg\langle H^v_i\mathscr{L}_{f^{(0)}}\left[\delta f^{(1)}\right] \bigg\rangle_{f_0} &=&\overbrace{-\frac{1}{2}\langle H^v_i\mathcal{A}_{kl}\rangle_{f^{(0)}}\left(\frac{\partial u_k}{\partial x_l}+\frac{\partial u_l}{\partial x_k}-\frac{2}{3}\frac{\partial u_s}{\partial x_s}\delta_{kl}\right)\nonumber -\frac{1}{\sqrt{\theta}}\left\langle H^v_i \mathcal{B}_{k}\right\rangle_{f^{(0)}}\left(\frac{\partial \theta}{\partial x_k}\right) }^{\mathcal{P}_i} 
\end{eqnarray}
on $\delta f^{(1)}$ \cite{Chapman1953}. 
By inserting the WE density in the constraint and exploiting the linearity of $\mathscr{L}_{f^{(0)}}[(.)]$, while keeping first-order terms, we get the system 
\begin{eqnarray}
\label{eq:Dlam_P}
\mathcal{D}_{ij}\lambda_j^{v(1)}&=&\mathcal{P}_i \ .
\end{eqnarray}
The above system can be solved for $\lambda^{v(1)}$ (since $\mathcal{D}$ is invertable) resulting in
\begin{eqnarray}
\label{eq:l-sol-NSF}
\lambda_i^{v(1)}&=&\mathcal{D}^{-1}_{ij}\mathcal{P}_j \ .
\end{eqnarray}
Furthermore, we can identify the viscosity and the heat conductivity. By definition, we get stresses and heat fluxes via
\begin{eqnarray}
\Sigma^{(1)}_{ij}&=&\theta\langle \mathcal{A}_{ij} H_k^v\rangle_{f^{(0)}}{\lambda^v_k}^{(1)} \\
q^{(1)}_i&=&\theta^{3/2}\langle \mathcal{B}_{i} H_k^v\rangle_{f^{(0)}}{\lambda^v_k}^{(1)} \ .
\end{eqnarray}
Now by inserting the explicit formulation of Lagrange multipliers Eq.~\eqref{eq:l-sol-NSF} we get
\begin{eqnarray}
\Sigma^{(1)}_{ij}&=&-\mu \left(\frac{\partial u_i}{\partial x_j}+\frac{\partial u_j}{\partial x_i}-\frac{2}{3}\frac{\partial u_s}{\partial x_s}\delta_{ij}\right) \\
q^{(1)}_i&=&-\kappa \frac{\partial \theta}{\partial x_i}
\end{eqnarray}
equipped with the viscosity
\begin{eqnarray}
\mu&=& \frac{\theta}{10}  \bigg(\langle \mathcal{A}_{kl}H_i^v\rangle_{f^{(0)}}\mathcal{D}^{-1}_{ij}\langle H_j^v\mathcal{A}_{kl}\rangle_{f^{(0)}} \bigg)
\label{eq:viscosity}
\end{eqnarray}
and the heat conductivity
\begin{eqnarray}
\kappa &=& \frac{\theta}{3}\bigg(\langle \mathcal{B}_{k}H_i^v\rangle_{f^{(0)}}\mathcal{D}^{-1}_{ij}\langle H_j^v\mathcal{B}_{k}\rangle_{f^{(0)}}\bigg) \ .
\label{eq:heat_cond}
\end{eqnarray}
 Therefore the first-order approximation in the limit of $\epsilon\to 0$, i.e. Eq.~\eqref{eq:moment-WE-NS}, gives us the NSF closure with the positive viscosity and heat conductivity (due to positive-definiteness of $\mathcal{D}$), similar to those obtained from MED (see equations (6.16) a and b in \cite{levermore1996moment}).  The resulting viscosity and heat conductivity are identical to those obtained from the Chapman-Esnkog expansion, once Maxwell molecular interaction is considered \cite{levermore1996moment}. However for a general interaction law, Eqs.~\eqref{eq:viscosity}-\eqref{eq:heat_cond} may deviate from Chapman-Enskog expressions \cite{Chapman1953}. 
\end{proof}
\section{Justification for Monte Carlo Scheme}
\label{proof:MC}
\noindent In the following we provide theoretical justification on the convergence of the stochastic scheme previously outlined. In particular, several properties have to be examined. Initially, we need to ensure the regularity of the introduced SDE system \eqref{eq:SDEs}. Next, we demonstrate that the proposed SDE system converges to a stationary solution that coincides with the WE closure. Finally, we show that the update in the Lagrange multiplier estimates $\tilde{\lambda}$ converges to the minimizer of Eq.~\eqref{eq:opt}. To improve readability, since the spatial position acts only as a parameter (and not a random variable), we omit the dependence on $x$ in our notation.\\ \ \\
Suppose we have given values for the Lagrange multipliers $\tilde{\lambda}$ which may or may not be optimal. It is more convenient, for what follows, to cast the SDE into the following form 
\begin{eqnarray}
\label{SDE-Z}
dZ_i &=&\tilde{\lambda}_j\partial_{z_i}H_j(z)\big\vert_{z=Z}\ dt-C_0\tilde{\alpha}\partial_{z_i}|v-w|^p\big\vert_{z=Z}\ dt+ \sqrt{2}\ \textrm{d}B_i
\end{eqnarray}
where $z=[v_1,v_2,v_3,w_1,w_2,w_3]^T$ and $\textrm{d}B$ is a six-dimensional Brownian process. We can identify the drift
\begin{eqnarray}
a_i(z)&=&\tilde{\lambda}_j\partial_{z_i}H_j(z)-C_0\tilde{\alpha}\partial_{z_i}|v-w|^p
\end{eqnarray}
and the diffusion $b$ as the identity matrix,
for this process. Furthermore, we make the following observations
\begin{enumerate}
\item The generator of this process on a smooth function $h(z)$ reads
\begin{eqnarray}
L[h(z)]&=&a_i\partial_{z_i}h+\partial^2_{z_iz_i}h \ .
\end{eqnarray}
\item The probability density $f_Z$ associated with $Z$ follows the Fokker-Planck equation
\begin{eqnarray}
\label{eq:FP-Z}
\partial_t f_Z&=&\partial_{z_i}\left(-a_if_Z\right)+\partial^2_{z_iz_i}f_Z \ .
\end{eqnarray}
\end{enumerate}
\begin{prop}
The SDE \eqref{SDE-Z} admits a unique solution global in time, provided bounded $\lambda^*$ and $\alpha>0$.
\end{prop}
\begin{proof}
The proof mainly follows \cite{khasminskii2011stochastic}. Due to the continuity of $a$ and constant diffusion, all we need to show is the stability of \eqref{SDE-Z}.  
Observe that the function
\begin{eqnarray}
s(z)&=&|v-w|^2
\end{eqnarray}
is positive and goes to $\infty$ as $|z|\to \infty$. Furthermore, we have  
\begin{eqnarray}
L[s(z)]&=&\left(\tilde{\lambda}_j\partial_{z_i}H_j(z)-C_0\tilde{\alpha}\partial_{z_i}|v-w|^p\right)\partial_{z_i}|v-w|^2+\partial^2_{z_iz_i}|v-w|^2 \ .
\end{eqnarray}
However since $-C_0\tilde{\alpha}|v-w|^p$ becomes dominant as $|z|\to \infty$, the same argument as in Proof \ref{proof:L1} can be used to show that $L[s]$ is bounded.  
Therefore according to Theorem 3.5 in \cite{khasminskii2011stochastic}, SDE \eqref{SDE-Z} has a unique solution for arbitrary $t$. \ 
\end{proof}
\begin{prop}
Suppose $f_Z$ follows the Fokker-Planck equation \eqref{eq:FP-Z}. As a result, $f_Z$ converges to
\begin{eqnarray}
h^*(z)&=&Z_h\exp\left(\tilde{\lambda}_iH_i-C_0\tilde{\alpha}|v-w|^p\right)
\end{eqnarray}
as $t\to\infty$, where $Z_h$ is the normalization factor. 
\end{prop}
\begin{proof}
We can easily check that $f_Z=h^*$ fulfills Eq.~\eqref{eq:FP-Z}, since $a_i=\partial_{z_i}(\log(h^*))$. In the following, we show that this solution will be attained as $t\to\infty$, independent of the initial condition. By defining the entropy distance 
\begin{eqnarray}
\bar{\mathcal{H}}&=&\langle \log (f_Z/h^*) \rangle_{f_Z},
\end{eqnarray}
we obtain 
\begin{eqnarray}
\partial_t \bar{\mathcal{H}}&=&-\left\langle \partial_{z_i}\log(f_Z/h^*)\partial_{z_i}\log(f_Z/h^*)\right\rangle_{f_Z} 
\end{eqnarray}
from Eq.~\eqref{eq:FP-Z}. 
The negativity of the right-hand side (the Fisher distance) guarantees the decay of $\bar{\mathcal{H}}$ to the minimum. However the minimum value of $\bar{\mathcal{H}}$ is attained once $f_Z=h^*$ and hence the SDE system converges to our closure with Lagrange multipliers $\tilde{\lambda}$.
\end{proof}
\begin{prop}
Let $\tilde{\lambda}(t)$ be updated by Eq.~\eqref{eq:WE_system_Lagrange_mult} consistent with the moment evolution given by \eqref{eq:rel}. Assuming all the involved coefficients are bounded, $\tilde{\lambda}(t)$ converges to the optimal $\lambda$, as the solution of the minimization problem \eqref{eq:opt}\ .
\end{prop}
\begin{proof}
Suppose
\begin{eqnarray}
Z_{\lambda^*}&=&\int_{\mathbb{R}^3 \times \mathbb{R}^3}\exp\bigg(\tilde{\lambda}_i H_i-\alpha C_0|v-w|^p \bigg) \   \text{dvdw} \ ,
\end{eqnarray}
and thus the minimization problem takes the form
\begin{eqnarray}
\lambda&=&\argminA_{\tilde{\lambda}\in\mathbb{R}^{2n}}\left\{\mathcal{F}({\tilde{\lambda}})\right\}, \ \ \ \text{where}\\  
\mathcal{F} &=& Z_{\tilde{\lambda}} -\tilde{\lambda}_i P_i.
\end{eqnarray}
We observe that the update imposed by the moment relaxation leads to an increment in $\tilde{\lambda}$ along the gradient of $\mathcal{F}(\tilde{\lambda})$.
More formally, the updates in $\tilde{\lambda}$ follow
\begin{eqnarray}
\partial_{t} \tilde{\lambda}_i &=&\frac{\partial \tilde{\lambda}_i}{\partial \tilde{P}_j}{\partial_t \tilde{P}_j}
\\
&=&\frac{\partial \tilde{\lambda}_i}{\partial \tilde{P}_j}\frac{\left(P_j-\tilde{P}_j\right)}{\tau} \ .
\end{eqnarray}
The first term on the right-hand-side can be expanded based on Eq.~\eqref{eq:ls}, leading to
\begin{eqnarray}
\frac{\partial \tilde{\lambda}_i}{\partial \tilde{P}_j}&=&-\tau^{-1}\left(\mathscr{A}\right)^{-1}_{ij}\ ,
\end{eqnarray}
where the right-hand-side is negative-definite due to positivity of $\mathscr{A}$. Next observe that the second term on the right-hand-side is proportionaal to the gradient of $\mathcal{F}$, since
\begin{eqnarray}
\frac{\partial \mathcal{F}(\tilde{\lambda})}{\partial \tilde{\lambda}_j}&=&\left(P_j-\tilde{P}_j\right).
\end{eqnarray}
Putting things together we have
\begin{eqnarray}
\partial_{t}\tilde{\lambda}_i&=&-\tau^{-2}\left(\mathscr{A}\right)^{-1}_{ij}\frac{\partial \mathcal{F}(\tilde{\lambda})}{\partial \tilde{\lambda}_j} \ .
\end{eqnarray}
Therefore the update in $\tilde{\lambda}$ is the product of a negative-definite matrix and gradient of the loss function $\mathcal{F}$. Hence the algorithm is of a gradient descent form, and therefore $\mathcal{F}$ decays due to the update of $\tilde{\lambda}$ \cite{wolfe1969convergence}, which yields the optimal solution $\tilde{\lambda}(t)\to\lambda$, as $t\to\infty$.
\end{proof}
\bibliographystyle{unsrt}
\bibliography{refs}

\end{document}